\newif\iffigs\figstrue
\documentclass[12pt]{article}
\iffigs
  \input epsf
\else
\fi

\batchmode
\newfont{\footscrfont}{rsfs10}
  \newfont{\footbbbfont}{msbm10}
  \newfont{\manfont}{manfnt}
\errorstopmode

\newif\ifscrf\scrftrue
\ifx\footscrfont\nullfont
  \scrffalse
\fi

\newif\ifamsf\amsftrue
\ifx\footbbbfont\nullfont
  \amsffalse
\fi

\def\ppnumber{\vbox{\baselineskip14pt\hbox{YITP-SB-00-69}
\hbox{hep-th/0010269}}}
\def\ppdate{October 2000}
\def\pplogo{\vbox{\kern-\headheight\kern -15pt
\halign{##&##\hfil\cr&{%\sc
\ppnumber}\cr\rule{0pt}{2.5ex}&\ppdate\cr}
}}

\makeatletter
\date{}
\def\dedicatory#1{\def\@date{\normalsize\it#1}}
\def\subjclass#1{\def\@thefnmark{}\@footnotetext{1991
    {\it Mathematics Subject Classification.} #1}}
\def\keywords#1{\def\@thefnmark{}\@footnotetext{
    {\it Key words and phrases.} #1}}

\def\ps@firstpage{\ps@empty \def\@oddhead{\hss\pplogo}%
  \let\@evenhead\@oddhead % in case an article starts on a left-hand page
}
\def\maketitle{\par
 \begingroup
 \def\thefootnote{\fnsymbol{footnote}}
 \def\@makefnmark{\hbox
 to 0pt{$^{\@thefnmark}$\hss}}
 \if@twocolumn
 \twocolumn[\@maketitle]
 \else \newpage
 \global\@topnum\z@ \@maketitle \fi\thispagestyle{firstpage}\@thanks
 \endgroup
 \setcounter{footnote}{0}
 \let\maketitle\relax
 \let\@maketitle\relax
 \gdef\@thanks{}\gdef\@author{}\gdef\@title{}\let\thanks\relax}

\def\abstract{\if@twocolumn
\section*{Abstract}
\else \small
\begin{center}
{\bf ABSTRACT}
\end{center}
\quotation
\fi}

\def\thebibliography#1{\section*{References\@mkboth
 {REFERENCES}{REFERENCES}}\small\list
 {[\arabic{enumi}]}{\settowidth\labelwidth{[#1]}\leftmargin\labelwidth
 \advance\leftmargin\labelsep
 \usecounter{enumi}}
 \def\newblock{\hskip .11em plus .33em minus .07em}
 \sloppy\clubpenalty4000\widowpenalty4000
 \sfcode`\.=1000\relax}

\newif\iffn\fnfalse

\@ifundefined{reset@font}{\let\reset@font\empty}{} %needed for ancient LaTeX
\long\def\@footnotetext#1{\insert\footins{\reset@font\footnotesize
    \interlinepenalty\interfootnotelinepenalty
    \splittopskip\footnotesep
    \splitmaxdepth \dp\strutbox \floatingpenalty \@MM
    \hsize\columnwidth \@parboxrestore
   \edef\@currentlabel{\csname p@footnote\endcsname\@thefnmark}\@makefntext
    {\rule{\z@}{\footnotesep}\ignorespaces
      \fntrue#1\fnfalse\strut}}}

\makeatother

\ifamsf
  \newfont{\bigbbbfont}{msbm10 scaled\magstep2}
  \newfont{\bbbfont}{msbm10 scaled\magstep1}  % msbm12 does not exist
  \newfont{\smallbbbfont}{msbm8}
  \newfont{\tinybbbfont}{msbm6}
  \newfont{\smallfootbbbfont}{msbm7}
  \newfont{\tinyfootbbbfont}{msbm5}
  \newfont{\biggthfont}{eufm10 scaled\magstep2}
  \newfont{\gthfont}{eufm10 scaled\magstep1}  % eufm12 does not exist
  \newfont{\smallgthfont}{eufm8}
  \newfont{\tinygthfont}{eufm6}
  \newfont{\footgthfont}{eufm10}
  \newfont{\smallfootgthfont}{eufm7}
  \newfont{\tinyfootgthfont}{eufm5}
\fi

\ifscrf
  \newfont{\scrfont}{rsfs10 scaled\magstep1}  % rsfs12 does not exist
  \newfont{\smallscrfont}{rsfs7}
  \newfont{\tinyscrfont}{rsfs7}
  \newfont{\smallfootscrfont}{rsfs7}
  \newfont{\tinyfootscrfont}{rsfs7}
\fi

\ifamsf
  \newcommand{\Bbb}[1]{\iffn
      \mathchoice{\mbox{\footbbbfont #1}}{\mbox{\footbbbfont #1}}
      {\mbox{\smallfootbbbfont #1}}{\mbox{\tinyfootbbbfont #1}}\else
      \mathchoice{\mbox{\bbbfont #1}}{\mbox{\bbbfont #1}}
      {\mbox{\smallbbbfont #1}}{\mbox{\tinybbbfont #1}}\fi}
  
\else
  \def\bigbbbfont{\bf}
  \def\Bbb{\bf}
  
\fi

\ifscrf
  \newcommand{\Scr}[1]{\iffn
    \mathchoice{\mbox{\footscrfont #1}}{\mbox{\footscrfont #1}}
    {\mbox{\smallfootscrfont #1}}{\mbox{\tinyfootscrfont #1}}\else
    \mathchoice{\mbox{\scrfont #1}}{\mbox{\scrfont #1}}
    {\mbox{\smallscrfont #1}}{\mbox{\tinyscrfont #1}}\fi}
\else
  \def\Scr{\cal}
\fi

\def\C{{\Bbb C}}

\def\Z{{\Bbb Z}}

\def\bearray{\begin{eqnarray}}
\def\eearray{\end{eqnarray}}
\def\bearraynn{\begin{eqnarray*}}
\def\eearraynn{\end{eqnarray*}}
\def\bfig{\begin{figure}}
\def\efig{\end{figure}}

\def\opeq#1{\advance\lineskip#1 \advance\baselineskip#1
        \advance\lineskiplimit#1}

\def\cM{{\Scr M}}

\def\cD{{\Scr D}}

\def\cMc{{\hfuzz=100cm\hbox to 0pt{$\;\overline{\phantom{X}}$}\cM}}
\def\barcD{{\hfuzz=100cm\hbox to 0pt{$\;\overline{\phantom{X}}$}\cD}}

\ifamsf

\else

\fi

\def\boldone{\relax{\rm 1\kern-.35em 1}}

\newtheorem{Proposition}{Proposition}[section]

\newtheorem{Theorem}{Theorem}[section]
\newtheorem{Lemma}{Lemma}[section]
\newtheorem{Corrolary}{Corrolary}[section]

\newcommand{\be}{\begin{equation}}
\newcommand{\ee}{\end{equation}}
\newcommand{\bea}{\begin{eqnarray}}
\newcommand{\eea}{\end{eqnarray}}

\newcommand{\bp}{\begin{Proposition}}
\newcommand{\ep}{\end{Proposition}}
\newcommand{\bt}{\begin{Theorem}}
\newcommand{\et}{\end{Theorem}}
\newcommand{\bl}{\begin{Lemma}}
\newcommand{\el}{\end{Lemma}}
\newcommand{\bc}{\begin{Corrolary}}
\newcommand{\ec}{\end{Corrolary}}
\newcommand{\nn}{\nonumber}

\usepackage{graphics}

\begin{document}

\title{On the structure of open-closed topological field theory in 
two-dimensions}

\author{C.~I.~Lazaroiu$^{*}$}

\date{}

\maketitle

\vbox{ \centerline{C.~N.~Yang Institute for Theoretical Physics} 
\centerline{SUNY at Stony Brook} 
\centerline{NY11794-3840, U.S.A.} 
\medskip 
\medskip
\bigskip }

\abstract{I discuss the general formalism of two-dimensional topological field 
theories defined on open-closed oriented Riemann surfaces, starting from an 
extension of Segal's geometric axioms.
Exploiting the
topological sewing constraints allows for the identification of 
the algebraic structure
governing such systems. I give a careful treatment of bulk-boundary and boundary-bulk
correspondences, which are responsible for the relation between the closed and open 
sectors. The fact that these correspondences need not be injective 
nor surjective has interesting implications for the problem of classifying 
`boundary conditions'. In particular, I give a clear geometric derivation of 
the (topological) 
boundary state formalism and point out some of its limitations. Finally, I formulate the problem 
of classifying (on-shell) 
boundary extensions of a given closed topological field theory in purely
algebraic terms and discuss their reducibility.}

\vskip .6in

$^*$ calin@insti.physics.sunysb.edu

\pagebreak

\tableofcontents

\pagebreak

\section{Introduction}

The central importance of open-closed strings
has become progressively clear since the discovery of D-branes. 
It is now generally accepted that a deeper understanding 
of open-closed string theory holds the key to deciphering not only D-brane 
dynamics but also some of the basic structures involved in non-perturbative proposals
for string theory such as M-theory. On the other hand, recent studies 
of open-closed string theory on Calabi-Yau 
manifolds hold the promise of providing new insight into the phenomena of 
mirror symmetry and topology change, as well as harmonizing 
the mathematical program of homological mirror symmetry \cite{Kontsevich} with modern 
developments in D-brane physics \cite{Ooguri,Douglas_quintic,Vafa, Vafa_mirror, Kachru,
SYZ}. 

It is somewhat surprising to notice that, in spite of its central importance, 
our understanding of open-closed string theory 
is quite incomplete when compared 
with the relatively well-developed framework available for the closed case. 
While considerable progress has been made in providing systematic constructions 
\cite{Fuchs, Recknagel, boundary, Zuber}, the current approach 
is largely based on the boundary state formalism \cite{Cardy}, which is sometimes
claimed to reduce most open-string questions to problems formulated
in the bulk. While 
this is certainly correct for some problems, 
this approach is in fact rather incomplete and cannot fully replace a systematic analysis of 
open-closed conformal/string theory in its native domain, namely through direct constructions 
motivated by the geometry of open-closed Riemann surfaces and
two-dimensional field theory dynamics. A clear approach to this problem seems 
especially important for studies of extended moduli spaces, which forms the core of the 
homological mirror symmetry program. 
Indeed, general points in the extended moduli space are not known to admit a
standard geometric description, and a clear definition of the systems under study is 
necessary in the absence of any intuitive considerations
\footnote{The reader used to the geometric approach via nonlinear sigma models will readily 
recognize simple realizations of our axiomatic constructions as they apply to that situation.
However, the main motivation of the present paper is to give a clear definition 
of open-closed topological field theory suitable for situations when a worldsheet path integral 
approach is either non-existent or unknown. In such cases, the meaning of `boundary conditions' is
unclear, and the classification of all boundary extensions of a given bulk
theory must be approached in abstract fashion.}.

The aim of this paper is to provide such an analysis for the 
simplified case of topological open-closed field theories in two dimensions. 
Beyond being technically simpler, such systems 
are bound to play a central role in current efforts to analyze D-brane dynamics 
in curved backgrounds. In particular, understanding their structure 
is crucial for studies of open-closed extensions of mirror symmetry.

By analogy with the closed case, open-closed topological strings can be built 
by coupling an open-closed topological field theory to topological gravity defined 
on open-closed Riemann surfaces (a generalization of the usual 
`closed' two-dimensional topological gravity of \cite{Witten_TG}). A detailed understanding 
requires a close look at each of these building blocks. In this paper I 
consider the first element only, namely the formalism of open-closed 
topological field theories.  These are distinguished from their 
gravitational counterpart by the fact that they do not contain a `dynamical' metric -
no integration over worldsheet metrics is necessary in order 
to achieve diffeomorphism invariance. I consider the abstract framework of such 
systems along the lines of \cite{Segal, Atiyah, Quinn} (see 
\cite{Dijkgraaf_leshouches} for a review). 
As in the closed case \cite{DVV,Dijkgraaf_notes,DVV_notes}, 
one can exploit the 
topology of bordered Riemann surfaces and 
the relevant axioms in order to encode 
all information about such theories 
into a finite set of characteristic (`structure') constants. These are 
subject to a set of conditions stemming from the 
topological sewing constraints, and I analyze these in order to 
extract the mathematical object they define. After making contact with the usual 
description in terms of correlators, I discuss how (a topological 
version of) the boundary state formalism can be recovered in this approach, and 
point out some of its conceptual limitations. I also give an abstract definition 
of a boundary extension of a topological bulk theory and shortly discuss the 
problem of irreducible versus reducible extensions. 
Finally, I discuss a rather obvious category-theoretic interpretation of 
boundary data and point out that this physically-motivated structure underlies 
recent work on  D-brane categories \cite{Douglas_categories}.

The formalism of the present paper is restricted to open-closed topological 
field theories on {\em oriented} Riemann surfaces. The  
unoriented case requires a slightly modified approach, which will not be 
discussed here. Some of the results derived below are 
probably 
familiar to topological field theory experts, though a clear, general and 
systematic 
derivation does not seem to have been given before. 
The expert reader may be interested in the 
detailed analysis of 
bulk-boundary and boundary-bulk maps and the topological 
version of 
the (generalized) Cardy constraint discussed in Section 4, as 
well as the discussion of reducibility 
and the category-theoretic interpretation  of Section 5. He may also 
be interested in our treatment of boundary-condition changing sectors.
The mathematical structure governing open-closed topological field 
theories is summarized in Subsection 4.8. I tried to make this
and Section 5 accessible to a mathematical audience, and to this end 
they collect some results derived in the rest of the paper in an attempt to 
make the presentation self-contained.

\section{Axiomatics}

\subsection{Surfaces, state spaces and products}

The framework of open-closed topological field theories in two dimensions
(`boundary' topological field theories) can be formulated through an extension 
of the geometric category approach of \cite{Segal,Atiyah} to the case of 
bordered Riemann surfaces. In this paper, we  restrict to the case of 
{\em oriented} strings, and hence consider oriented Riemann surfaces only.
Since we allow for general boundary conditions (i.e. we define our theory in 
the presence of D-branes), each open string boundary will carry a label 
(decoration) $a$, which indicates the associated boundary sector
\footnote{When a topological sigma model description is 
available, the indices $a$ label various boundary conditions/choices of 
Chan-Paton data.}.
Our Riemann surfaces $\Sigma$ carry two types of boundaries. First, one has
`closed' and `open' string boundaries. The former are oriented circles $C$, while the 
latter are oriented segments $I$. 
The `open string boundaries' $I$ carry boundary sector 
labels $a,b$ at their ends, which we indicate by writing $C_a$ or $I_{ba}$ (in the latter 
case, the convention is that $I$ is oriented from $a$ to $b$).
Second, one has `boundary sector' boundaries, which are oriented open or closed 
curves $\gamma_a$ carrying a single label $a$. These are those 
bounding curves of $\Sigma$ on which the 
`boundary conditions' are imposed.
Since we deal with a topological field theory,
we consider all objects up to orientation-preserving diffeomorphisms, which 
is to say that their parameterizations do not matter. 

We shall declare a string boundary $C$ or $I_{ba}$ to be `incoming' if its orientation 
agrees with that of $\Sigma$ and `outgoing' otherwise. Physically, such boundaries 
are associated with incoming/outgoing strings. 
A topological field theory `living' on such surfaces defines a bulk state 
space ${\cal H}$ (obtained through quantization on the infinite cylinder) and a 
collection of boundary state spaces ${\cal H}_{ba}$ (obtained through 
quantization on an infinite strip carrying boundary conditions $a$ and 
$b$). 
Our convention for the latter is that ${\cal H}_{ba}$
corresponds to the state space of the oriented 
open string stretching from $a$ to $b$ (in this order). The bulk and boundary 
state spaces ${\cal H},{\cal H}_{ba}$ are $\Z_2$-graded:
\bea
{\cal H}&=&{\cal H}^0\oplus {\cal H}^1~~,\\
{\cal H}_{ba}&=&{\cal H}_{ba}^0\oplus{\cal H}_{ba}^1~~,
\eea
where the $\Z_2$-degree of a state can be identified with its Grassmann parity. Hence 
states belonging to ${\cal H}^0,{\cal H}^0_{ba}$ are `bosonic' (and, when 
such a description is available, generated 
by Grassmann-even worldsheet fields), while states 
in ${\cal H}^1,{\cal H}^1_{ba}$ are `fermionic' (and generated by Grassmann-odd 
fields)
\footnote{For topological sigma models, the $\Z_2$-grading is induced
by a $\Z$-grading associated with the worldsheet $U(1)$ charge: 
states of even $U(1)$ charge are Grassmann even and states of odd $U(1)$ charge are Grassmann
odd. A general model does not possess a worldsheet $U(1)$ symmetry, since it need not be 
obtained by twisting an $N=2$ superconformal field theory. However, the $\Z_2$-degree 
is always defined.}. If a state $\phi$ has pure degree $d$, we shall 
write $deg\phi=|\phi|=d$. We will sometimes use ket notation $\phi=|\phi\rangle$.

A surface $\Sigma$, together with an enumeration of its incoming and outgoing string 
boundary components, defines 
a map $\Phi_\Sigma$ (called the associated {\bf product}) between the  
incoming and outgoing state spaces. These are defined through
${\cal H}_{in}:=\otimes_{i=1}^{m}{{\cal H}_{\Gamma^{in}_i}}$, 
${\cal H}_{out}:=\otimes_{j=1}^n{{\cal H}_{\Gamma^{out}_j}}$, where $\Gamma^{in}_i$, 
$\Gamma^{out}_j$ are the (enumerated) incoming and outgoing 
string boundary components of $\Sigma$. 
Let us recall the path integral definition for completeness
\footnote{In practice, one often obtains a realization of 
these axioms through cohomological field theories (such as the A/B models); in this case, 
the two-dimensional metric enters as an explicit parameter and becomes 
irrelevant only after taking the cohomology of a nilpotent operator $Q$. 
The path integral derivation of sewing constraints does not directly apply to such models, 
through it is easy to show that they satisfy our axioms {\em on-shell}, 
i.e. {\em after} taking $Q$-cohomology. For brevity, we illustrate the 
path-integral origin only for the simpler case of strict 
(i.e. off-shell metric-independent) topological field theories.}. In the path integral 
formalism, one associates a configuration space with each string boundary of $\Sigma$. 
In our situation, one has a bulk configuration space $V$ (the space of configurations 
of worldsheet fields restricted to a string bounding circle $C$) and open configuration 
spaces $V_{ba}$ (the space of field configurations on a string bounding interval, 
subject to the boundary conditions labeled by $a$ and $b$ at its two ends).
Both bulk and boundary configuration spaces are (infinite-dimensional) 
supermanifolds\footnote{Because we study topological field theories,  
Grassmann-odd worldsheet/boundary 
configurations are {\em not} spinors from the worldsheet
point of view. This is important when considering sewing operations which 
produce nontrivial closed curves, in which case the path integral gives 
objects of the type $Tr((-1)^F\{\dots\})$. The factor $(-1)^F$ is due to 
the fact that odd configurations are always periodic along such cycles.
This is familiar from the case of twisted sigma models, where the 
G-odd fields are related to Ramond sector fermions of the untwisted model.}
. Next, one 
defines ${\cal H}, {\cal H}_{ba}$ 
as the spaces of functionals over configuration spaces (functions defined on the 
supermanifolds $V$ and $V_{ba}$). Their $\Z_2$-grading is induced by  
Taylor expansion with respect to odd coordinates on $V,V_{ba}$.

Given an enumeration of incoming/outgoing boundaries of $\Sigma$, define the incoming and outgoing 
configuration spaces by $V_{in}=\times_{i=1}^{m}{V_{\Gamma^{in}_i}}$  and 
$V_{out}=\times_{j=1}^{n}{V_{\Gamma^{out}_j}}$. The enumeration of incoming/outgoing 
string boundaries does matter in these definitions, since the configuration spaces 
$V$, $V_{ba}$ contain Grassmann-odd elements.
Picking $\phi^{in}\in V_{in}$, 
and $\phi^{out}\in V_{out}$, we next consider the (Euclidean) 
path integral over field configurations
$\phi$ on $\Sigma$ subject to the boundary conditions $\phi|_{\Gamma^{in}_i}=
\phi^{in}_i$ and $\phi|_{\Gamma^{out}_j}=\phi^{out}_j$ on the string boundaries, 
and to the boundary conditions indexed by the label $a$ on each other boundary $\gamma_a$:
\be
K_{\Sigma}(\phi_{out},\phi_{in})=\int_{\phi|_{\Gamma^{out}_j}=
\phi^{out}_j,\phi|_{\Gamma^{in}_i}=\phi^{in}_i}{{\cal D}[\phi]e^{-S[\phi]}}~~.
\ee

This gives a function $K_\Sigma$ defined on $V_{out}\times V_{in}$. 
It allows us to define the map $\Phi_{\Sigma}$ from ${\cal H}_{in}=
\otimes_{\Gamma_{in}}
{{\cal H}_{\Gamma_{in}}}$ to 
${\cal H}_{out}=
\otimes_{\Gamma_{out}}{{\cal H}_{\Gamma_{out}}}$ as follows. For each incoming state 
$\eta\in {\cal H}_{in}$, 
we define the associated outgoing state $\eta'=\Phi_{\Sigma}(\eta)$ to be the
function(al) on $V_{out}$ given by the following equation:
\be
\eta'(\phi_{out})=\int{{\cal D}[\phi_{in}]K_\Sigma(\phi_{out},\phi_{in})
\eta(\phi_{in})}~~.
\ee
Here ${\cal D}[\phi_{in}]$ is the path integral measure on boundary configurations.

In the particular case when all boundaries of $\Sigma$ are incoming, the outgoing space 
associated
with $\Sigma$ is ${\cal H}_{out}=\C$ and the map $\Phi_\Sigma$ is a complex-valued 
linear 
functional defined on the incoming space. This is 
the {\em correlator} defined by $\Sigma$, and will also be denoted by 
$\langle \dots\rangle_\Sigma$. 

\subsection{Axioms}

\subsubsection{Degree}
Topological products are subject to a {\bf degree axiom}, which requires that 
all products with a single output are maps of 
degree zero, while the maps with two inputs and no output (the topological 
metrics, see below) have definite (but model-dependent) degree. 

\subsubsection{Sewing}
The topological surfaces $\Sigma$ can be composed by sewing at their closed or 
open string boundary components. Sewing is allowed only between two closed string boundaries 
or two open string boundaries, and the orientations and endpoint 
labels of the sewn boundaries must match. 
Since we deal with a topological field theory, parameterizations at the 
boundaries do not matter, so there is no twist-sewing operation. 
Sewing defines an associative composition on the collection of 
topological open-closed Riemann surfaces, which endows it with the structure of a category. In this 
category, the objects are direct products of 
closed and open string boundaries, i.e. oriented 
topological circles and oriented segments with endpoint decorations. 
The morphisms are the Riemann surfaces 
themselves--mapping incoming into outgoing boundary components, 
while sewing gives the morphism compositions. 
Since the objects of our category are given by direct products, 
their components are ordered. Thus the Riemann surfaces 
connecting them are endowed with
distinguished enumerations of the incoming and outgoing string boundaries. 
What we have is a generalization of 
(the topological version of) Segal's geometric category \cite{Segal}. 

\hskip 1.0 in
\begin{center} 
\scalebox{1.5}{\epsfxsize=6cm \epsfbox{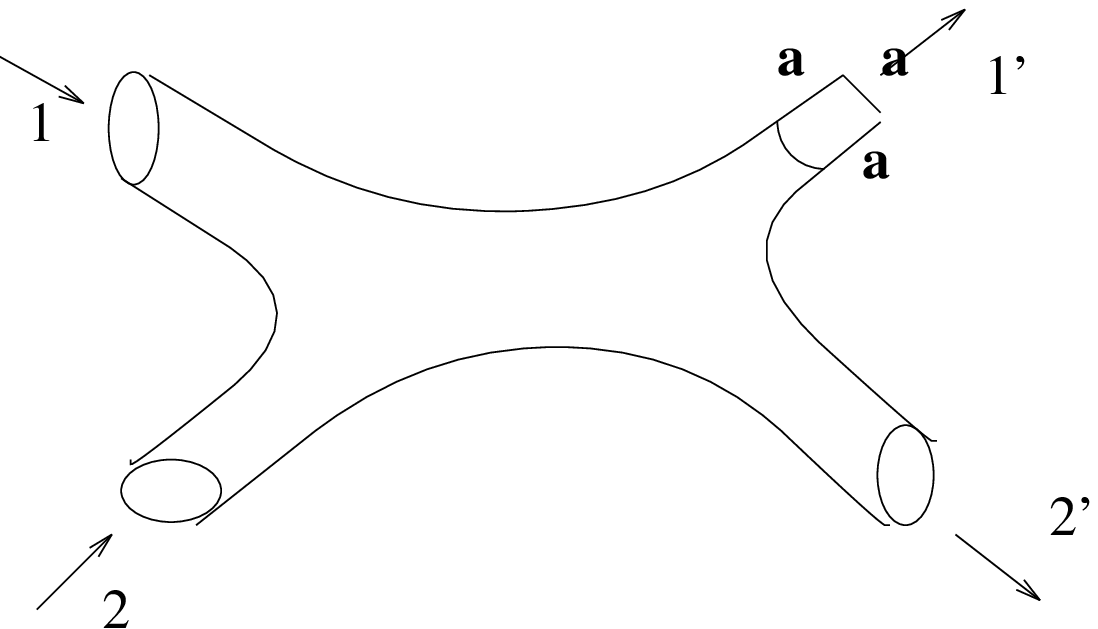}}
\end{center}
\begin{center} 
Figure  1. {\footnotesize A typical open-closed Riemann surface}
\end{center}

The {\bf sewing axiom} is the 
requirement that the correspondence $\Sigma\rightarrow \Phi_\Sigma$
be a functor from this geometric category to the linear category defined by 
tensor products of the 
spaces ${\cal H}, {\cal H}_{ba}$ together with linear maps between such products.
This requires that sewing of two Riemann surfaces $\Sigma$ and $\Sigma'$
corresponds to composition of the associated maps $\Phi_\Sigma$ and 
$\Phi_{\Sigma'}$:
\be
\Sigma\infty\Sigma'\rightarrow \Phi_\Sigma\circ \Phi_{\Sigma'}~~.
\ee
The sewing axiom can be 
`derived' from elementary properties of the path integral in the standard manner. 

\subsubsection{Permutation symmetry (equivariance)}
One also requires that the correspondence $\Sigma\rightarrow \Phi_\Sigma$ be 
{\bf graded equivariant} with respect to arbitrary permutations of closed string boundaries and 
cyclic permutations of open string boundaries. For closed string boundaries, 
such permutations correspond to the associated action on the tensor product components of ${\cal H}_{in}$ and 
${\cal H}_{out}$, and the map $\Phi_{\Sigma_{perm}}$ determined by the 
`permuted' surface should be related to $\Phi_\Sigma$ by composing with these 
linear operations at its ends. The latter permutations act 
with signs dictated by the degree of the permuted elements (permuting two 
G-odd states gives a minus sign etc.). 
For open boundaries, the 
condition is imposed on diagrams whose open string boundaries
are all incoming or all outgoing. This corresponds to graded-cyclic 
symmetry of open amplitudes. Note that only cyclic permutations 
are allowed, even for the case when all open string boundaries carry the 
same label\footnote{This fact is familiar in boundary conformal field
theory. In that case, open amplitudes are invariant under  
arbitrary permutations of the boundary insertions (assuming that all such
boundaries carry the same label) only if such insertions are 
`mutually local' (see  the second reference of \cite{Recknagel}). This happens, for example,
for those open boundary correlators which can be continued to the bulk.} 
(figure 2). 

\hskip 1.0 in
\begin{center} 
\label{permutations_fig}
\scalebox{1.0}{\epsfxsize=6.5cm \epsfbox{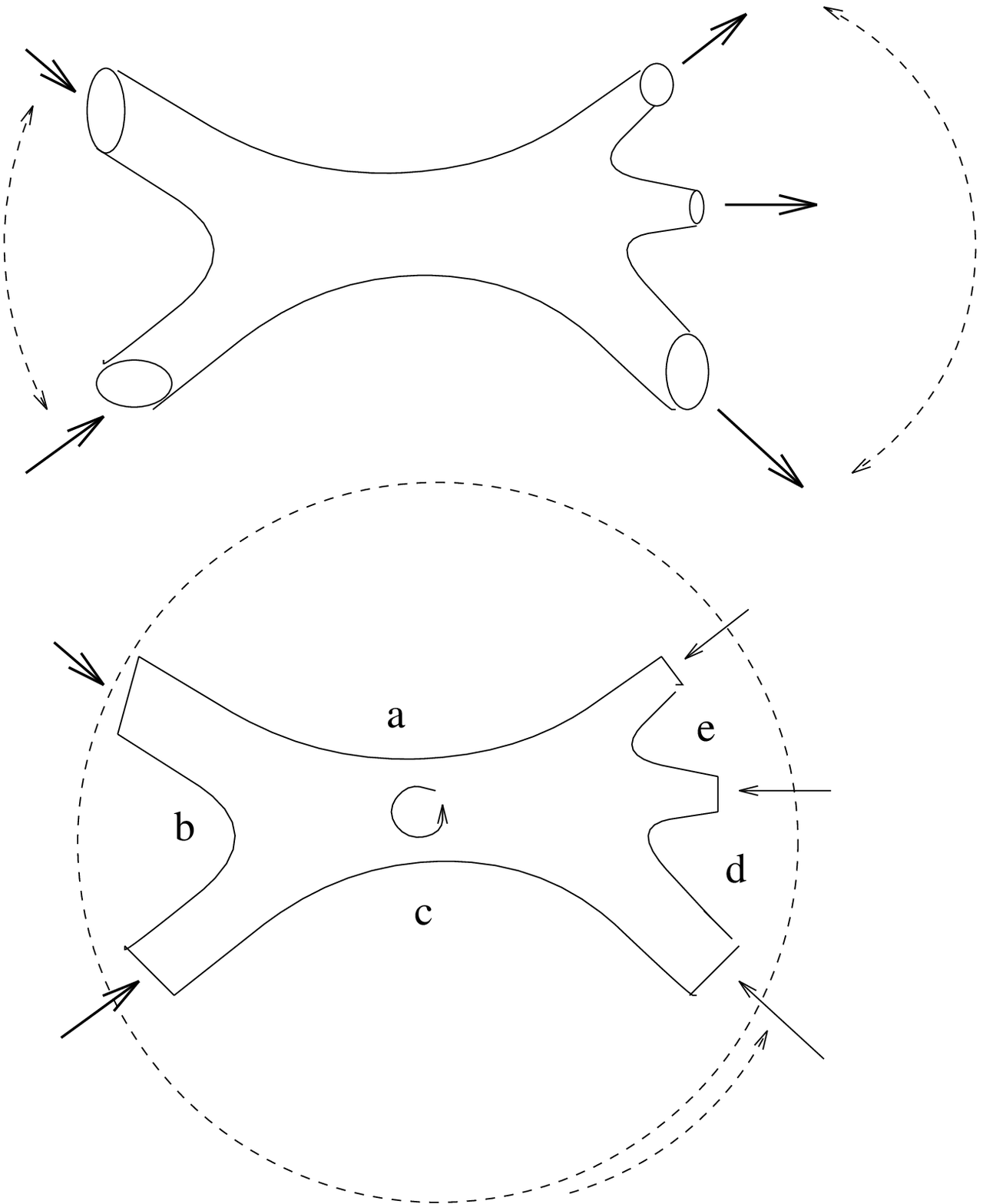}}
\end{center}
\begin{center} 
Figure 2. {\footnotesize Permutations allowed in the equivariance axiom.
For closed boundaries, any permutation is allowed among incoming or 
outgoing data. For open boundaries, equivariance requires cyclic symmetry 
of amplitudes. This condition refers to diagrams having only incoming 
or only outgoing open boundaries with the topology of a segment. Only 
cyclic permutations are allowed in the second diagram, even for the case 
$a=b=c=d=e$.}
\end{center}

\subsubsection{Normalization}

Finally, we have to impose a {\bf normalization constraint}. This  
requires that the linear maps defined by the surfaces in figure 3 are the 
identity operators of the corresponding state spaces, and encodes triviality of  
topological propagators\footnote{Cohomological field theories 
satisfy our axioms only {\em after} taking BRST 
cohomology. For such models, the bulk and boundary Hamiltonians $H$ and
$H_{ba}$ are usually BRST 
exact, and they induce trivial propagators in BRST cohomology. However, off-shell 
propagation in such models is nontrivial.}.

\hskip 1.0 in
\begin{center} 
\scalebox{1.5}{\epsfxsize=6cm \epsfbox{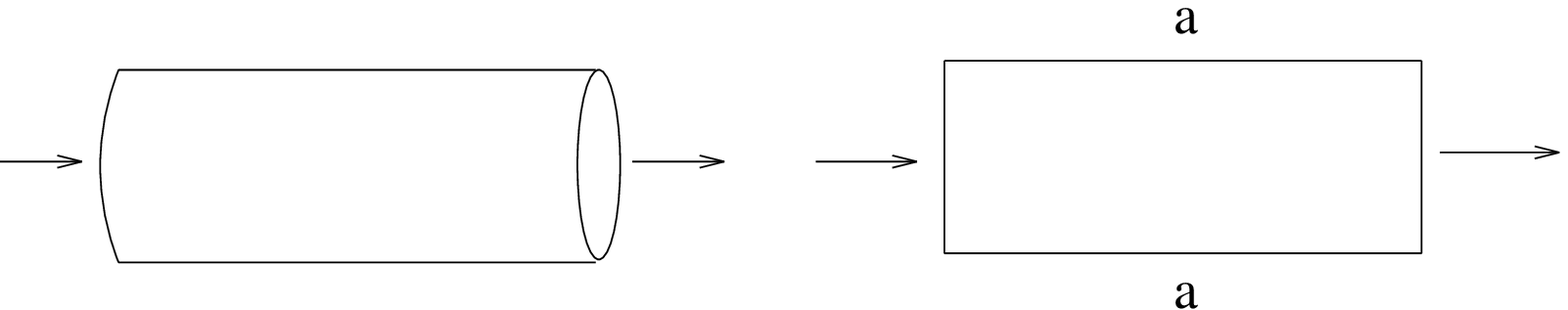}}
\end{center}
\begin{center} 
Figure  3. {\footnotesize The surfaces entering the normalization axiom.}
\end{center}

\section{Basic products, boundary vacua, metrics and 
traces and their first properties}

Any oriented open-closed Riemann surface (with any choice of orientation of its string boundaries) 
can be obtained 
by sewing some combination of the five basic surfaces shown in figure 4.
This is the analogue of the well-known pants decomposition of closed Riemann 
surfaces. For want of a better name, we shall call the three surfaces shown in 
figure 4 (a, b, c) 
by the names of closed pants, open pants and open-closed conduits. Beyond these, 
we also need two exceptional surfaces, namely the cylinder and the half-strip
with certain string boundary orientations, 
which are shown in figure 4 (d,e).

\hskip 1.0 in
\begin{center} 
\scalebox{1.5}{\epsfxsize=6cm \epsfbox{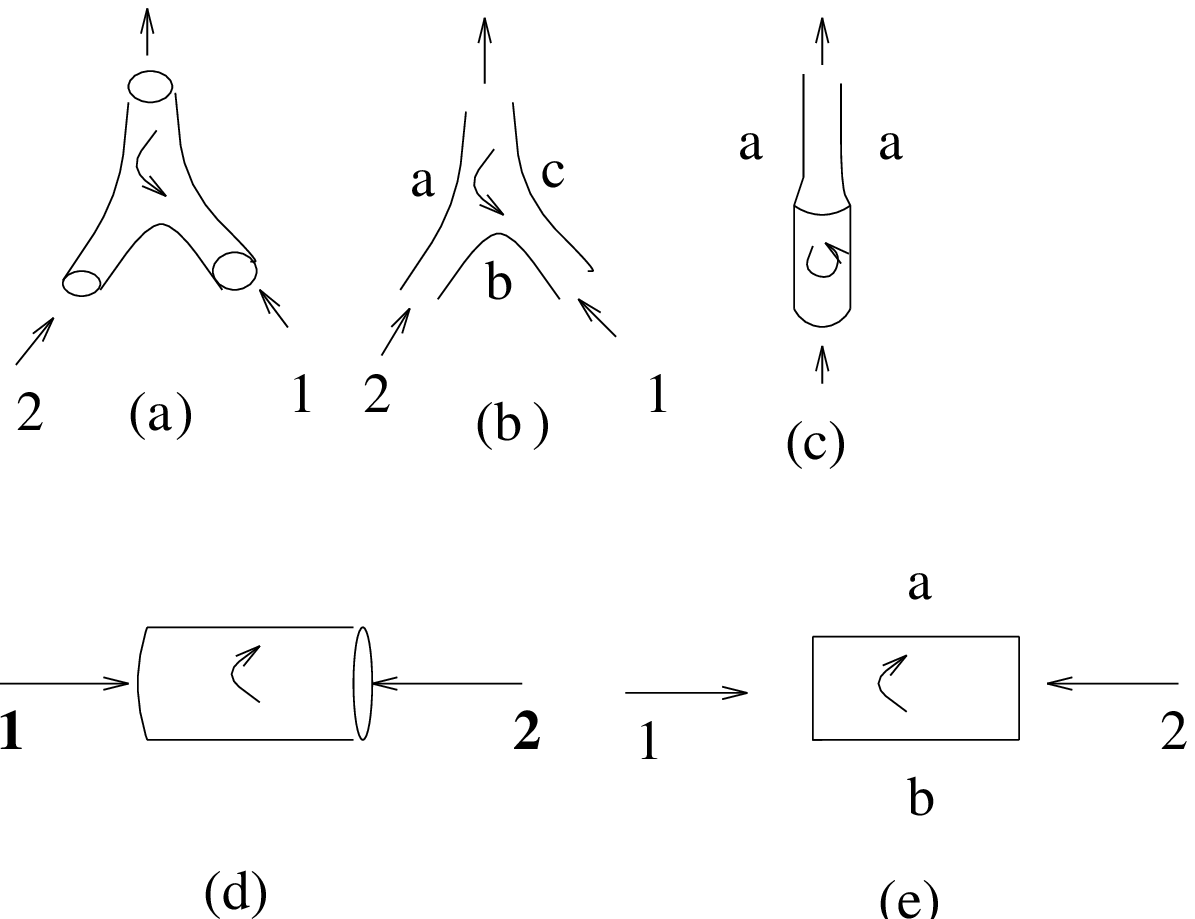}}
\end{center}
\begin{center} 
Figure  4. {\footnotesize The basic open-closed Riemann surfaces are considered 
with the indicated boundary orientations. Note that sewing with one of the 
surfaces $(d,e)$ allows us to revert the orientation of any outer leg of the 
surfaces $(a,b,c)$. }
\end{center}

The sewing axiom allows us to decompose an arbitrary topological product into 
the products defined by the basic surfaces. Hence the entire 
information about our theory is encoded by some basic data, which we 
now consider in turn. 

\subsection{The closed (bulk) product}

This is the degree zero bilinear product $C:{\cal H}\times {\cal H}\rightarrow {\cal H}$ 
defined by the surface in figure 4 (a). Since closed Riemann surfaces 
form a closed subclass under sewing, we can immediately identify this with the 
basic product of the associated closed topological field theory. It defines 
a bulk state-operator correspondence $g$, as follows. To each state 
$|\phi\rangle\in {\cal H}$, 
we associate the operator ${\hat \phi}:=g(|\phi\rangle)$ from ${\cal H}$ to ${\cal H}$ given by:
\be
{\hat \phi}(|\phi'\rangle):=C(|\phi\rangle,|\phi'\rangle)~~.
\ee
This parallels the bulk 
state-operator correspondence of closed conformal field theories, 
albeit in a simplified fashion. 

Since the vector space ${\cal H}$ is typically finite-dimensional, 
we can choose a finite 
basis $|\phi_i\rangle$ 
and define the coefficients $C_{ij}^k$ through the expansions:
\be
C(|\phi_i\rangle,|\phi_j\rangle)=\sum_{k}{C_{ij}^k|\phi_k\rangle}~~.
\ee
These are the well-known bulk structure constants familiar from 
closed topological field theory. 
Via the state-operator 
correspondence, the product $C$ corresponds to usual composition:
\be
g(|\phi_i\rangle)\circ g(|\phi_j\rangle)=g(C(|\phi_i\rangle,|\phi_j\rangle))~~.
\ee
This follows from the definition of $g$, by using associativity of the bulk 
product $C$, to be discussed below.

\subsection{The open (boundary) product}

The open pants of figure 4(b) define a degree zero bilinear product 
$B(cba):{\cal H}_{cb}\times {\cal H}_{ba}
\rightarrow {\cal H}_{ca}$. We introduce a boundary 
state-operator correspondence $g(cba)$ which associates to each state $|\psi\rangle$ of 
${\cal H}_{cb}$ an operator ${\hat \psi}^{(a)}:=g(cba)(|\psi\rangle)$ 
from 
${\cal H}_{ba}$ to ${\cal H}_{ca}$: 
\be
{\hat \psi}^{(a)}(|\psi'\rangle):=B(cba)(|\psi\rangle,|\psi'\rangle)\in {\cal H}_{ca}~~,
\ee
where $|\psi'\rangle\in {\cal H}_{ba}$. Choosing bases 
$|\psi_\alpha^{ba}\rangle$ for all spaces ${\cal H}_{ba}$, we can 
define {\em boundary structure constants} $B_{\beta\alpha}^\gamma(cba)$ via:
\be
B(cba)(|\psi_\beta^{cb}\rangle, |\psi_\alpha^{ba}\rangle)=
\sum_{\gamma}{B_{\beta\alpha}^\gamma(cba)|\psi_\gamma^{ca}\rangle}~~.
\ee
Associativity of the boundary product (discussed below) implies that 
the boundary state-operator correspondence takes the boundary product into usual 
operator compositions:
\be
g(cbe)(\psi_2)\circ g(bae)(\psi_1)=g(cae)(B(cba)(\psi_2,\psi_1))~~,
\ee
for $\psi_1\in {\cal H}_{ba}$ and $\psi_2\in {\cal H}_{cb}$.

As we shall see in more detail below, the role of the `diagonal' spaces 
${\cal H}_a:={\cal H}_{aa}$ is slightly different from that of 
the `off-diagonal' spaces ${\cal H}_{ba}$ with $b\neq a$. Following standard
terminology, the operators $g(aae)$ defined by states
$\psi^{a}_\alpha:=\psi^{aa}_\alpha$ will be called 
`boundary operators in the sector $a$', while the operators $g(bae)$ defined by $\psi^{ba}_\alpha$
with $b\neq a$ will be called `topological boundary condition changing 
operators'. They are the topological counterparts of CFT operators bearing the 
same names.

\subsection{The bulk-boundary maps}

The surface of figure 4 (c) defines  degree zero {\em bulk-boundary} maps\footnote{In 
a topological sigma model, these are realized through restriction of the bulk 
fields to the string boundary
components of the Riemann surface $\Sigma$. No such interpretation exists for systems 
which do not admit a sigma model description.},
which we denote by $e(a):{\cal H}\rightarrow {\cal H}_{a}$. These  
take the closed (bulk) state space ${\cal H}$ into each `diagonal' boundary state 
space ${\cal H}_a={\cal H}_{aa}$. There is generally no such map
into the `off-diagonal' spaces ${\cal H}_{ba}$ ($b\neq a$). Expressing 
this map in the bases $\phi_i$ and $\psi_\alpha^a$ 
allows us to define {\em bulk-boundary coefficients} $e_i^\alpha(a)$ for 
each boundary label $a$:
\be
e(a)(|\phi_i\rangle)=\sum_{\alpha}{e_i^\alpha(a)|\psi_\alpha^a\rangle}~~.
\ee
The bulk and boundary state--operator correspondences translate this  
into the {\em bulk-boundary expansions}. These are the maps 
$E(ab)=g(aab)\circ e(a)\circ g^{-1}$ between the bulk and boundary operator spaces. We have:
\be
E(ab)({\hat \phi}_i)=\sum_{\alpha}{e_i^\alpha(a)({\hat \psi}_\alpha^a)^{(b)}}~~.
\ee
This relation is the topological counterpart of the bulk-boundary expansion in boundary 
conformal field theories \cite{Cardy}. In the conformal case, it is 
usually written without explicit indication of (the conformal analogue of) the
map $E(ab)$ or of the label $b$.
Following the same convention, we could rewrite it in the more familiar
form:
\be
\label{formal}
{\hat \phi}_i=\sum_{\alpha}{e_i^\alpha(a){\hat \psi}_\alpha^a}~~.
\ee
However,  the maps $e(a),E(ab)$ {\em need not be 
injective nor surjective}
\footnote{Explicit examples of non-injectivity/non-surjectivity are provided by Calabi-Yau 
sigma models.}, hence care must be used when interpreting formal 
relations such as (\ref{formal}).

\subsection{Topological metrics}

The last two surfaces of figure 4
define complex bilinear maps $\eta:{\cal H}\times {\cal H}\rightarrow \C$ and
$\rho(ab):{\cal H}_{ab}\times {\cal H}_{ba}\rightarrow \C$, 
the bulk and boundary {\em topological metrics}. The equivariance axiom 
shows that these have the graded symmetry properties:
\bea
\eta(\phi_1,\phi_2)&=&(-1)^{|\phi_1||\phi_2|}\eta(\phi_2,\phi_1)~~\\
\rho(ab)(\psi_1,\psi_2)&=&(-1)^{|\psi_1||\psi_2|}\rho(ba)(\psi_2,\psi_1)~~.
\eea

The metrics are invariant with respect to the bulk and boundary products, respectively, as can be seen 
from figure 5:
\bea
\label{invariance}
\eta(C(\phi_1,\phi_2),\phi_3)~~~&=&\eta(\phi_1,C(\phi_2,\phi_3))~~\\
\rho(ac)(B(abc)(\psi^{ab}_1,\psi^{bc}_2),\psi^{ca}_3)
&=&\rho(ab)(\psi^{ab}_1,B(bca)(\psi^{bc}_2,\psi^{ca}_3))~~.
\eea
\hskip 0.5in
\begin{center} 
\scalebox{1.0}{\epsfxsize=7cm \epsfbox{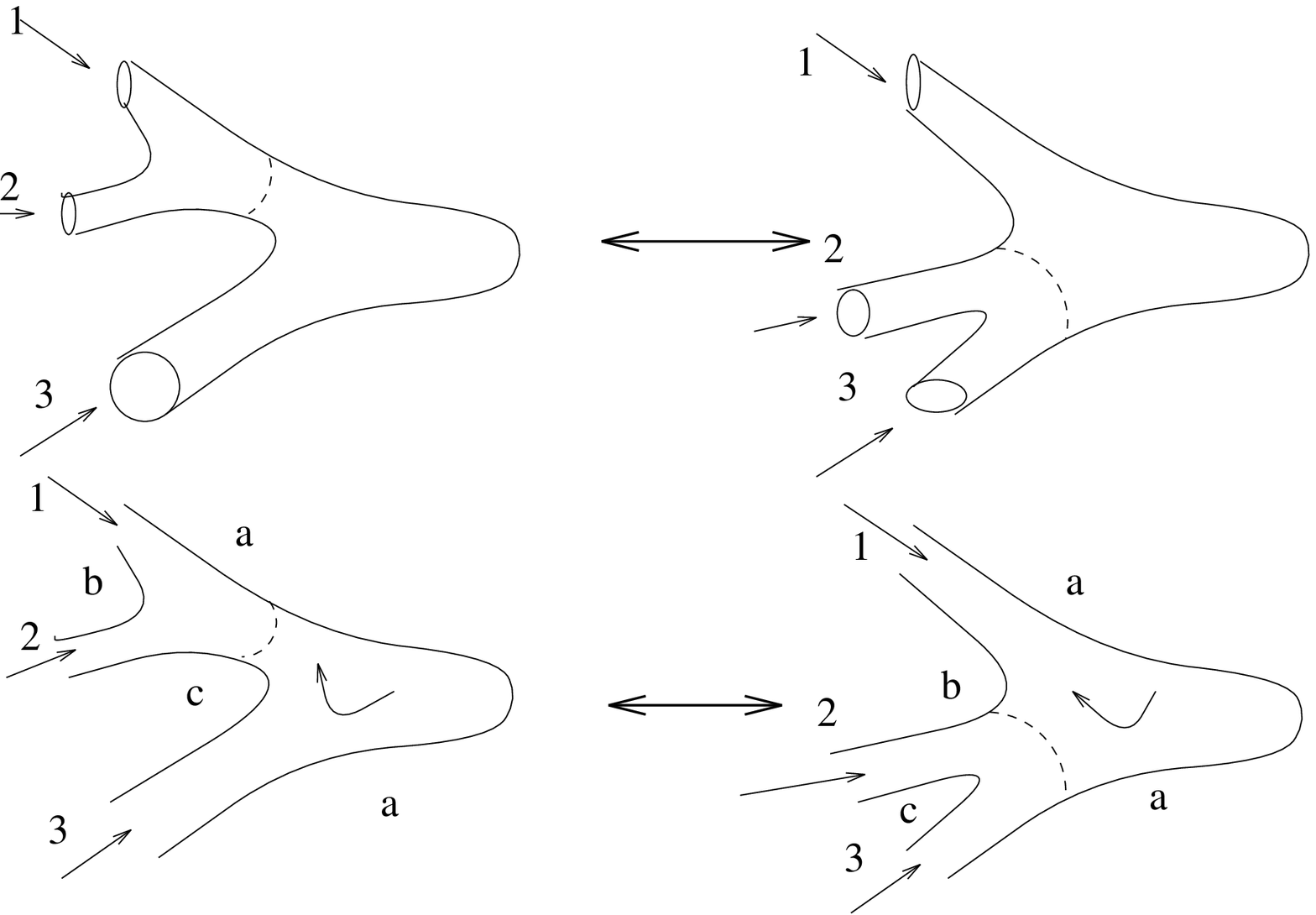}}
\end{center}
\begin{center} 
Figure  5. {\footnotesize Invariance of the topological metrics with respect to the bulk and 
boundary products.}
\end{center}

Moreover, figure 6  shows that the metrics are non-degenerate as bilinear forms
\footnote{We assume that our state spaces are all finite dimensional, which is the usual case
in practice.}:
\bea
\label{nondegeneracy}
\eta(\phi_1,\phi_2)~~=0~{\rm~for~all~}~\phi_1~&\Rightarrow& \phi_2=0~~\\
\rho(ab)(\psi^{ab}_1,\psi^{ba}_2)=0~{\rm~for~all~}~\psi_1^{ab}&\Rightarrow& \psi^{ba}_2=0~~\\
\rho(ab)(\psi^{ab}_1,\psi^{ba}_2)=0~{\rm~for~all~}~\psi_2^{ba}
&\Rightarrow& \psi^{ab}_1=0~~.
\eea

\begin{center} 
\scalebox{1.5}{\epsfxsize=6cm \epsfbox{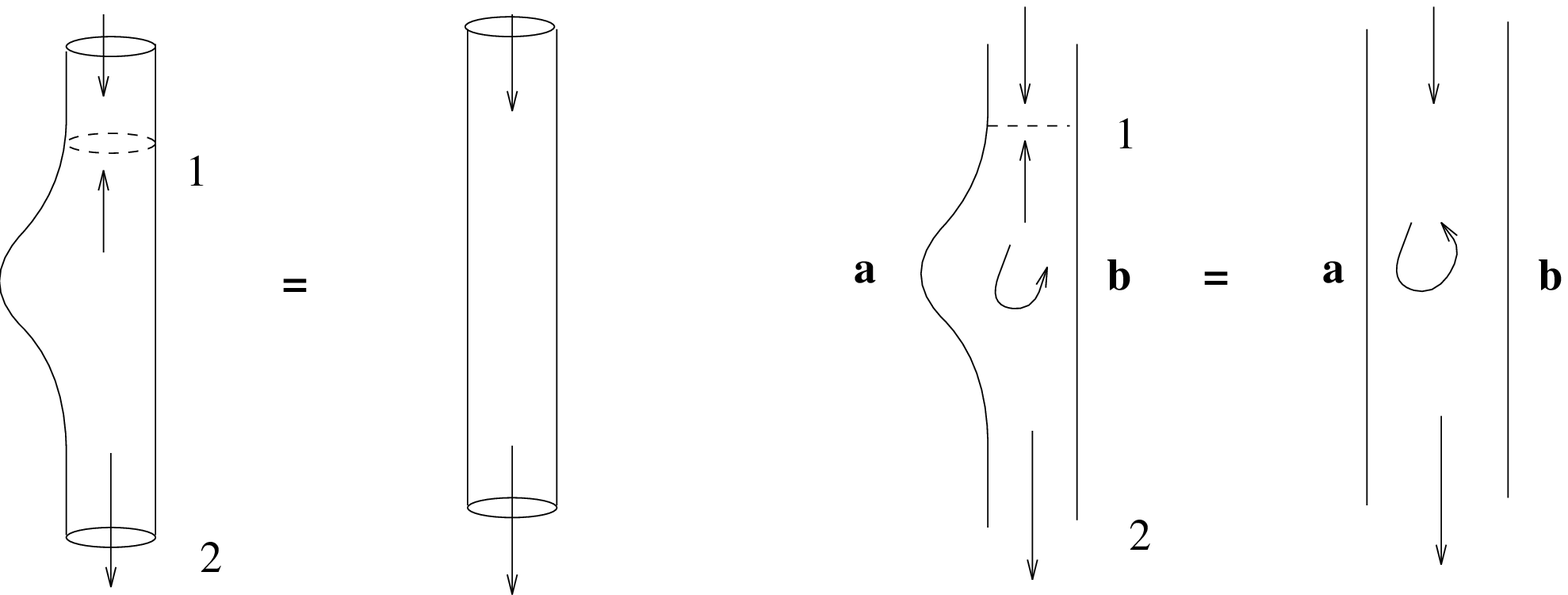}}
\end{center}
\begin{center} 
Figure  6. {\footnotesize Graphical construction of `inverses' for the topological metrics. The 
two-legged surfaces below the cuts define states $q\in {\cal H}\otimes {\cal H}$ and 
$p(ab)\in {\cal H}_{ab}\otimes{\cal H}_{ba}$. The figure shows that 
$(\eta\otimes id_{\cal H})(\phi\otimes q)=q$ and
$(\rho(ab)\otimes id_{{\cal H}_{ba}})(\psi^{ab}\otimes p)=\psi^{ab}$. 
This implies that the topological metrics are non-degenerate.}
\end{center}

Hence the topological  metrics allow us to identify ${\cal H}^*$ with 
${\cal H}$ and ${\cal H}^*_{ab}$ with ${\cal H}_{ba}$, where $^*$ indicates the 
linear dual of a vector space. Defining metric coefficients by:
\bea
\eta_{ij}&:=&\eta(|\phi_i\rangle,|\phi_j\rangle)~~\\
\rho_{\alpha\beta}(ab)&:=&\rho(ab)(|\psi_\alpha^{ab}\rangle,|\psi_\beta^{ba}\rangle)~~,
\eea
we introduce functionals $\langle \phi_i|\in {\cal H}^*$ and 
$\langle \psi_\alpha^{ab}|\in {\cal H}_{ab}^*$ through the conditions:
\bea
\langle \phi_i|(|\phi_j\rangle)&:=&\langle \phi_i|\phi_j\rangle=\eta_{ij}~~\label{duals1}\\
\langle \psi_\alpha^{ab}|(|\psi_{\beta}^{ab}\rangle)&:=&
\langle \psi_\alpha^{ab}|\psi_{\beta}^{ab}\rangle=\rho_{\alpha\beta}(ab)~~.\label{duals2}
\eea
We have the completeness relation:
\be
\label{completeness_bulk}
\sum_{i,j}{|\phi_i\rangle\eta^{ij}\langle\phi_j|}=id_{{\cal H}}~~,
\ee
where $\eta^{ij}$ is the matrix inverse to $\eta_{ij}$.

For the boundary sector, we define a map
$F(ab):{\cal H}_{ab}\rightarrow {\cal H}_{ba}$ through:
\be
\label{completeness_boundary}
F(ab):=\sum_{\alpha,\beta}{|\psi_\alpha^{ba}\rangle\rho^{\alpha\beta}(ab)
\langle\psi_\beta^{ab}|}~~,
\ee
with $\rho^{\alpha\beta}(ab)$ the inverse of $\rho_{\alpha\beta}(ab)$.
This takes $|\psi_\alpha^{ab}\rangle$ into $|\psi_\alpha^{ba}\rangle$ 
for all $\alpha$, and thus gives an isomorphism between ${\cal H}_{ab}$ and 
${\cal H}_{ba}$. Identifying these two spaces via the isomorphism $F$, 
we can treat them as identical, in which case $F$ can be viewed as the 
identity operator of the space ${\cal H}_{ab}\approx{\cal H}_{ba}$. Then 
(\ref{completeness_boundary}) can be understood as the completeness relation for 
the basis $|\psi_\alpha^{ab}\rangle\approx |\psi_\alpha^{ba}\rangle$ in this 
vector space.

Note that the bulk and boundary topological metrics are not related in any simple 
fashion. In particular, the boundary topological metric is {\em not} the 
`boundary restriction' of the bulk metric\footnote{In a topological sigma model, 
the bulk-boundary map $e$ is typically given by `restriction to the boundary'. 
However, the action of the model will generally contain a nonzero boundary term, 
such as a boundary coupling to a gauge connection. The boundary metric is given 
by a path integral on the strip (upper half plane punctured at the origin), and 
hence depends on the boundary action. The bulk metric is given by a path integral 
on  a cylinder (complex plane punctured at the origin), and depends only on the 
bulk action.}. That is, one need {\em not} have 
$\rho(a)(e(a)(\phi_1),e(b)(\phi_2))=\eta(\phi_1,\phi_2)$, as can be seen from the 
geometry of the associated Riemann surfaces\footnote{We use the notation $\rho(a):=\rho(aa)$ 
for the diagonal boundary sectors.}.

\subsection{Reduction to correlators}

Given a surface $\Sigma$ (without a choice of orientation for its boundary components $\Gamma_i$), 
one can use it to define various products $\Phi_{\Sigma,O}$ associated to the possible orientations
$O$ of these components. 
A `canonical' choice is to consider incoming boundaries only, in which case one
obtains the correlator $\langle \dots\rangle_\Sigma$. This can be related to the other products defined
by $\Sigma$ with the help of the topological metrics. For the example, let 
$\langle \dots \rangle_\Sigma$ be the correlator 
defined by the surface and boundary orientations shown in figure $7$. 
Then cutting very close to the outgoing boundary gives:
\be
\langle\psi_1^{ab}\psi_2^{bc}\psi_3^{ca}\rangle_\Sigma=\rho(ac)(m(\psi_1^{ab},\psi_2^{bc}),\psi_3^{ca})~~, 
\ee
where $m$ is the product defined by the three-pronged surface determined by the cut.
\hskip 0.5in
\begin{center} 
\scalebox{1.0}{\epsfxsize=5cm \epsfbox{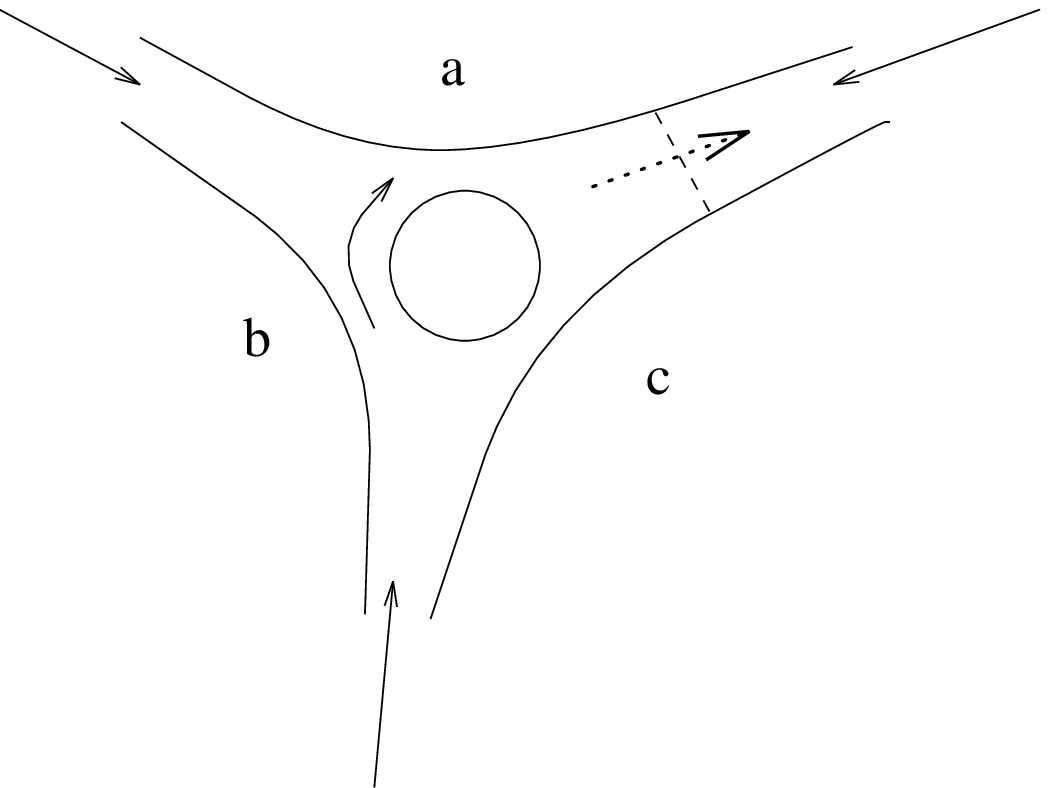}}
\end{center}
\begin{center} 
Figure  7. {\footnotesize Relating topological correlators to other products.}
\end{center}

Due to non-degeneracy of the topological metrics, we conclude that all products $\Phi_\Sigma$ 
can be determined from the knowledge of correlators. Similarly, they can all be determined
from knowledge of the topological metrics and of  products with a single output.

Using the state-operator correspondence, we can identify correlators 
with the topological vevs of the associated operator products
\footnote{This involves the observation that $g(|\phi\rangle)|0\rangle
=C(|\phi\rangle,|0\rangle)=|\phi\rangle$
and $g(baa)(|\psi^{ba}\rangle)|0_a\rangle=B(baa)(|\psi^{ba}\rangle,|0_a\rangle)=|\psi^{ba}\rangle$,
where $|0\rangle$ and $|0_a\rangle$ are the topological vacua discussed below, and the definition 
of dual states given in eqs. (\ref{duals1}) and (\ref{duals2}).}.
This recovers the usual formalism.

\subsection{Units}

The surfaces of figure 8 define degree zero linear maps 
from the field $\C$ of complex numbers into the spaces ${\cal H}$ and 
${\cal H}_a$. Evaluating these maps at the complex identity $1\in \C$
defines special degree zero states which we denote by $|0\rangle\in {\cal H}$ 
and $|0_a\rangle\in {\cal H}_a$. These states play the role of 
`topological vacua' in their respective spaces, as can be seen by considering 
the surfaces shown in figure 9 below. This figure shows that 
$C(|0\rangle,|\phi_j\rangle)=|\phi_j\rangle, C(|\phi_i\rangle,|0\rangle)=
|\phi_i\rangle$ and $B(aab)(|0_a
\rangle,
|\psi_\beta^{ab}\rangle)=|\psi_\beta^{ab}\rangle,~
B(abb)(|\psi_\alpha^{ab}\rangle, |0_b\rangle)=|\psi_\alpha^{ab}\rangle$, i.e.:
\be
C_{0j}^k=\delta^k_j~,~C_{i0}^k=\delta^k_i{\rm and}~~
B_{0\beta}^\gamma(aab)=\delta^\gamma_\beta~,~B_{\alpha0}^\gamma(abb)=\delta^\gamma_\alpha~~.
\ee
It follows that the states 
$|0\rangle$ (in ${\cal H}$) and $|0_a\rangle$ (in ${\cal H}_a$) associated with 
these surfaces are neutral elements
(units) with respect to the bulk and boundary operator products. Note that there 
is no natural definition of a topological vacuum in a boundary condition 
changing sector ${\cal H}_{ab}$ with $a\neq b$. 

\hskip 1.0 in
\begin{center} 
\scalebox{1.5}{\epsfxsize=6cm \epsfbox{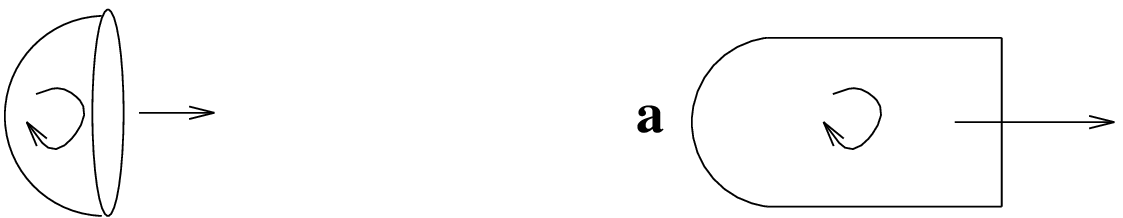}}
\end{center}
\begin{center} 
Figure  8. {\footnotesize Defining surfaces for the topological vacua.
Each of these surfaces contains a single string boundary, namely the 
circle/segment on their right. The boundary topological vacua 
arise from a path integral with boundary condition labeled $a$ on the non-string 
boundary. This gives a functional on the space of open string states supported
on the string boundary, which is the segment $I_{aa}$ to the right. }
\end{center}

\hskip 1.0 in
\begin{center} 
\scalebox{1.5}{\epsfxsize=6cm \epsfbox{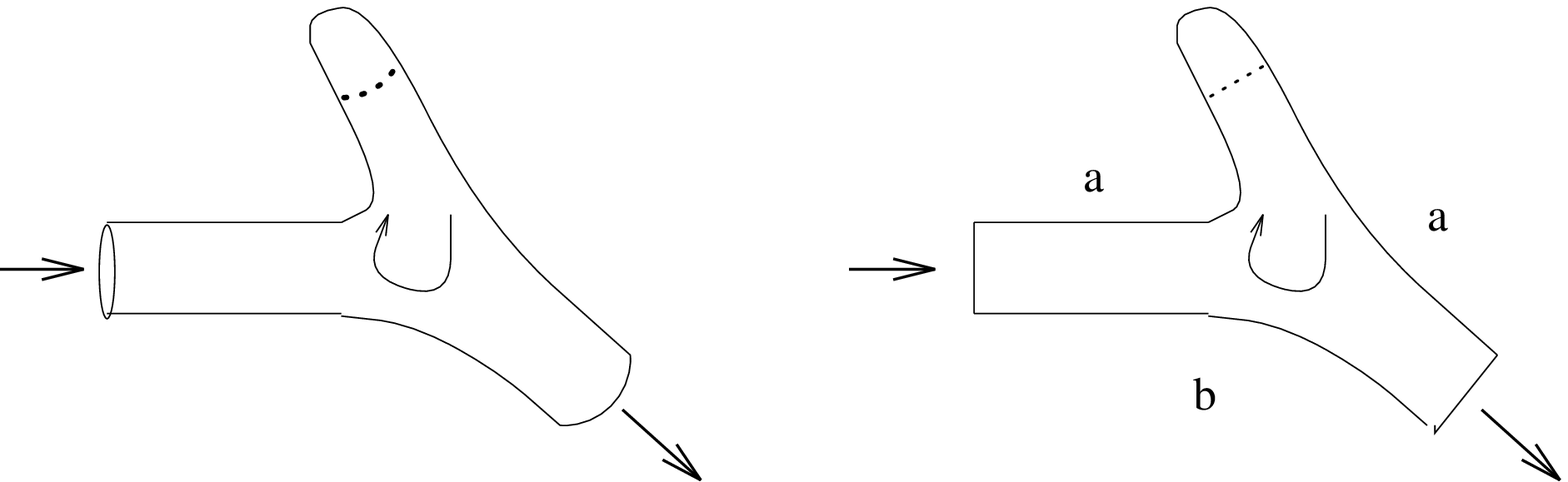}}
\end{center}
\begin{center} 
Figure  9. {\footnotesize The topological vacua are units for the basic topological products.}
\end{center}

Investigation of figure 10 shows that the boundary vacua are 
related to the bulk vacuum through the maps $e(a)$:
\be
\label{bb_vacua}
e(a)|0\rangle=|0_a\rangle
\ee

\hskip 1.0 in
\begin{center} 
\scalebox{1.5}{\epsfxsize=6cm \epsfbox{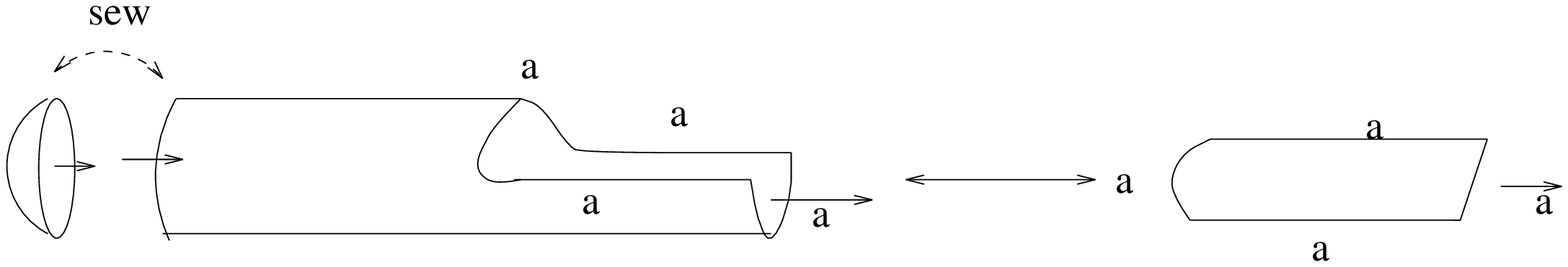}}
\end{center}
\begin{center} 
Figure  10. {\footnotesize Relation between the boundary and bulk 
topological vacua. Sewing the cap to the `conduit' of figure 4 (c) 
gives a surface which is topologically equivalent with the half strip.}
\end{center}

\subsection{Topological traces (=topological one-point functions)}

Considering the surfaces of figure $8$ with the opposite string 
boundary orientations gives linear maps
$Tr$, $Tr_{a}$ from the spaces ${\cal H},{\cal H}_{a}$ to the field of complex numbers. 
Figure 11 shows that the bulk and (diagonal) boundary topological 
metrics can be expressed in terms of products 
and traces:
\bea
\eta(\phi_1,\phi_2)~~&=&Tr(C(\phi_1,\phi_2))~~\label{metriC_trace1}\\
\rho(ab)(\psi_1^{ab},\psi_2^{ba})&=&Tr_a(B(aba)(\psi_1^{ab},\psi_2^{ba}))~~.\label{metriC_trace2}
\eea
\hskip 0.4in
\begin{center} 
\scalebox{1.0}{\epsfxsize=6cm \epsfbox{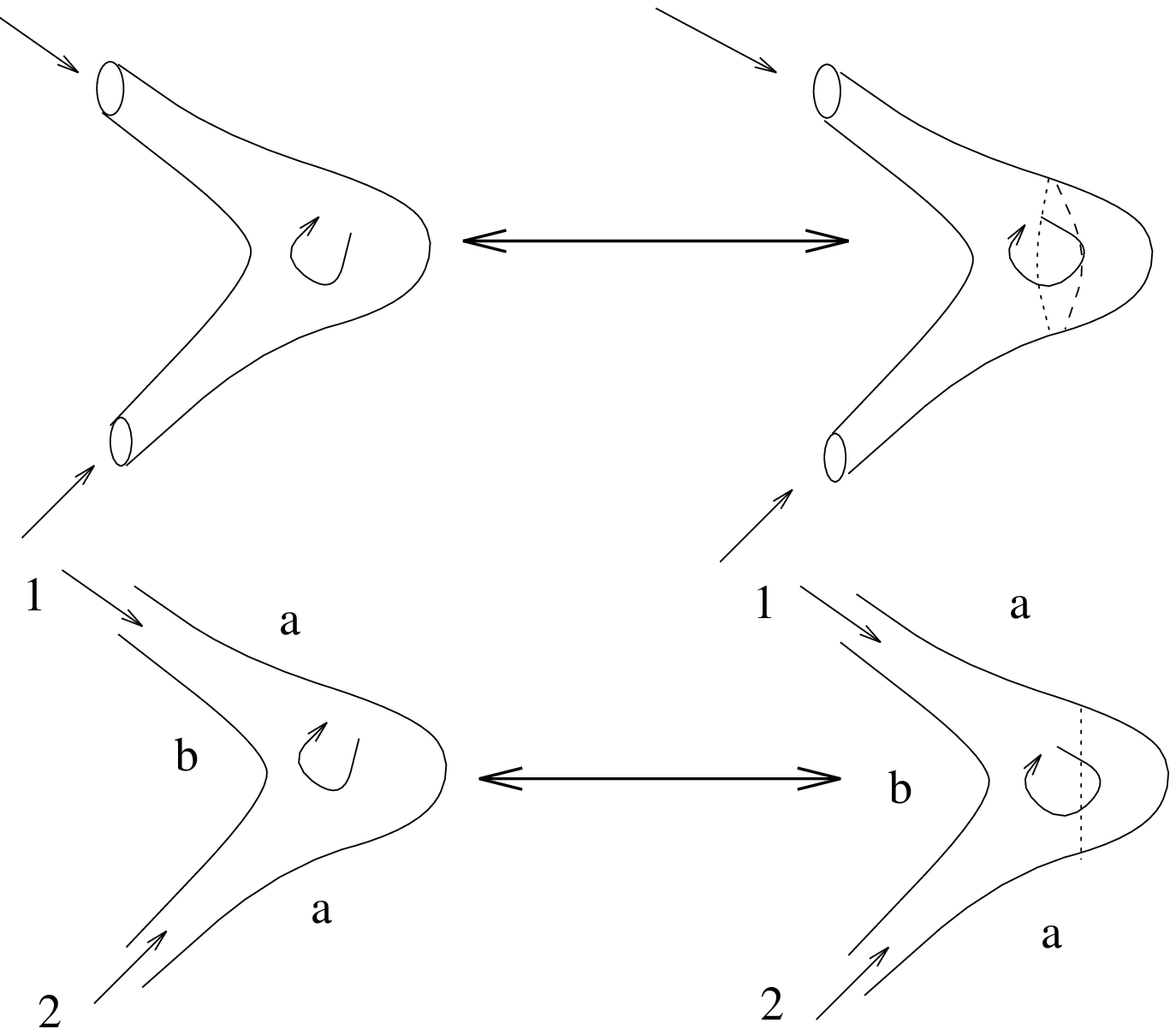}}
\end{center}
\begin{center} 
Figure  11. {\footnotesize The relation between topological traces and metrics.}
\end{center}

\noindent In particular, we have:
\bea
\eta(|0\rangle,\phi)&=&Tr(\phi)~~\\
\rho(ab)(|0_a\rangle,\psi^{a})&=&Tr_a(\psi^{a})~~.
\eea
Hence $Tr$, $Tr_a$
are the linear functionals on ${\cal H}, {\cal H}_{a}$
dual to the topological vacua $|0\rangle$, $|0_a\rangle$ 
with respect to the topological metrics $\eta,\rho(a)$.

\section{Consequences of the sewing constraints}

We saw that the entire information about an open-closed topological field theory 
is encoded by the three classes of products $C,B,e$ and the topological 
metrics $\eta$ and $\rho$. As in boundary conformal field theory, 
consistency of topological amplitudes under different 
decompositions of the same Riemann surface into the basic surfaces of figure 4
imposes constraints on this data. The sewing constraints in the conformal case 
have been analyzed in detail in \cite{Lewellen}, and since we deal with a 
topological  field theory (which is, in particular, conformally invariant), 
we can apply some of those results. The main observation of 
\cite{Lewellen} is that all sewing constraints are satisfied provided 
that the five basic conditions described in figure 12 are obeyed. The first 
two conditions are the basic sewing constraints of the closed
case (bulk crossing duality and bulk modular covariance), while the remaining 
conditions (shown in figure 12 (c,d,e,f)) encode boundary, 
open-open-closed and closed-open-open crossing duality and a supplementary constraint 
relating the bulk and boundary sector. 
We shall analyze the consequences of these conditions on our basic data. 
In fact, it turns out that the constraint of figure 12(b) is not 
required in the topological case (it reduces to a tautology for topological field 
theories). We have included it in our discussion since we want to stress similarity 
with the analysis of \cite{Lewellen}.

\

\

\hskip 2.0 in
\begin{center} 
\scalebox{1.0}{\epsfxsize=7cm \epsfbox{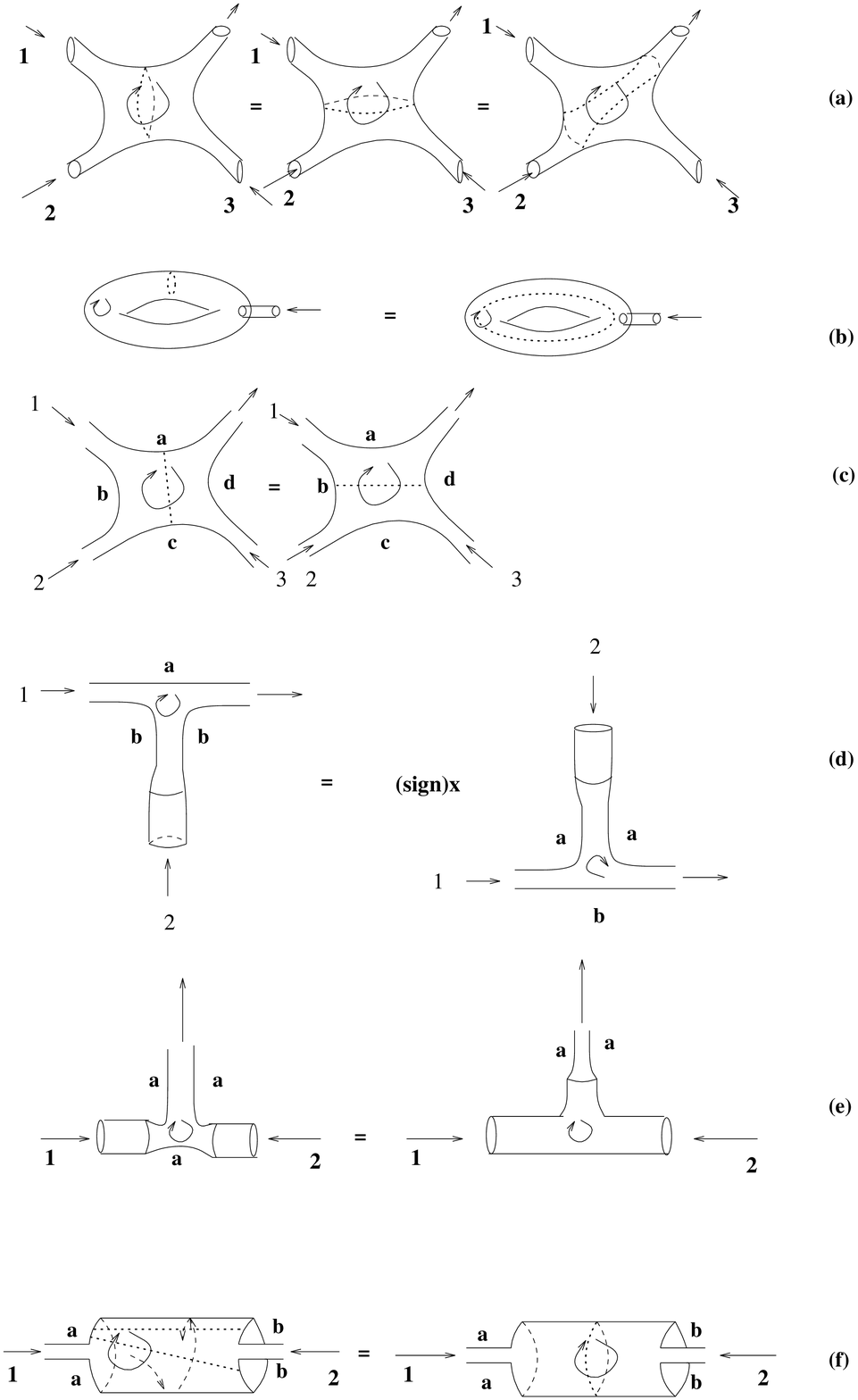}}
\end{center}
\begin{center} 
Figure  12. {\footnotesize Graphical depiction of the sewing constraints. The constraint (b) 
is void for topological field theories, as explained later in this section.}
\end{center}

\

\subsection{Bulk crossing symmetry}

As in closed topological field theory, the sewing constraint described by 
figure 12 (a) amounts to the statement that the product $C$ must be 
associative. On the other hand, the equivariance axiom requires that this 
product is super-commutative:
\be
C(\phi_1,\phi_2)=(-1)^{|\phi_1||\phi_2|}C(\phi_2,\phi_1)~~.
\ee

Since $|0\rangle$ is a unit, we conclude that $C$ defines a structure 
of associative, super-commutative ring with a unit on the bulk state space 
${\cal H}$. Since the product is bilinear, this ring is in fact an algebra 
over the field of complex numbers. Moreover, we know that 
the bulk topological metric is graded-symmetric and invariant with respect to the product $C$. 
An associative, super-commutative $\C$-algebra with unit, 
endowed with a non-degenerate, graded-symmetric, 
invariant bilinear form is called a  
{\em Frobenius superalgebra} \cite{Dubrovin, DVV, Dijkgraaf_notes, Voronov}. Thus 
we recover the well-known fact 
that the bulk data $({\cal H},B,\eta)$ define a Frobenius superalgebra.

\subsection{Boundary crossing symmetry}

The constraint described by figure 12 (c) can be written:
\be
B(acd)(B(abc)(\psi_1,\psi_2),\psi_3)=B(abd)(\psi_1,B(bcd)(\psi_2,\psi_3))~~,
\ee
which is a `decorated' associativity condition.
On the other hand, equivariance requires that the triple correlator on the 
decorated disk is (graded) cyclically symmetric:
\bea
\langle \psi^{ab}_1 \psi^{bc}_2 \psi^{ca}_3\rangle_{abc}&=&
(-1)^{|\psi^{ab}_1|(|\psi^{bc}_2|+|\psi^{ca}_3|)} 
\langle \psi^{bc}_2 \psi^{ca}_3 \psi^{ab}_1\rangle_{bca}=\nn\\
&=&(-1)^{|\psi^{ca}_3|(|\psi^{ab}_1|+|\psi^{bc}_2|)} 
\langle \psi^{ca}_3 \psi^{ab}_1 \psi^{bc}_2\rangle_{cab}
\eea
Hence the boundary product is cyclic
in the following sense:
\bea
\label{cyclicity}
\rho(ac)(B(abc)(\psi_1,\psi_2),\psi_3)&=&
(-1)^{|\psi_1|(|\psi_2|+|\psi_3|)}
 \rho(ba)(B(bca)(\psi_2,\psi_3),\psi_1)=\nn\\
&=&(-1)^{|\psi_3|(|\psi_1|+|\psi_2|)}
\rho(cb)(B(cab)(\psi_3,\psi_1),\psi_2)~~.
\eea

\subsection{Consequences for traces}

Associativity of the bulk and boundary products implies that relations
(\ref{metriC_trace1}) and (\ref{metriC_trace2}) 
generalize to all tree-level bulk and boundary amplitudes:
\bea
\langle \phi_1,\dots,\phi_n\rangle_{0,0}&=&Tr(\phi_1 \dots \phi_n)~~\\
\langle \psi_1^{a_1a_2},\dots,\psi_{n-1}^{a_{n-1}a_n},\psi_n^{a_na_1}
\rangle^{a_1\dots a_n}_{0,1}&=&Tr_{a_1}(
\psi_1^{a_1a_2}\dots \psi_n^{a_na_1})~~,
\eea
where juxtaposition in the right hand side stands for the bulk or boundary product.
It is clear from (\ref{metriC_trace1}) and (\ref{metriC_trace2}) 
that the traces and topological metrics
determine each other, provided that the bulk and boundary products are known. Hence one can view 
the traces as `derived' concepts, if we treat the topological metrics as 
`fundamental'. The properties of the topological metrics imply that the traces 
are graded cyclic:
\bea
Tr(\phi_1\phi_2\phi_3)~~~~&=&(-1)^{|\phi_1|(|\phi_2|+|\phi_3|)}Tr(\phi_2\phi_3\phi_1)~~\\
Tr_a(\psi^{ab}_1\psi^{bc}_2\psi^{ca}_3)&=&(-1)^{|\psi^{ab}_1|(|\psi^{bc}_2|+|\psi^{ca}_3|)}
Tr(\psi^{bc}_2\psi^{ca}_3\psi^{ab}_1)~~,
\eea
which in particular justifies their name. Cyclicity of the traces also follows by applying 
the equivariance axiom to three-point correlators. In particular, we see that the bulk trace 
obeys a stronger constraint, which allows us to commute arbitrary entries:
\be
Tr(\phi_1\phi_2)=(-1)^{|\phi_1||\phi_2|}Tr(\phi_2\phi_1)~~.
\ee
This generally is not allowed for the boundary traces $Tr_a$. Instead, we have
the relation:
\be
Tr_a(\psi_1^{ab}\psi_2^{ba})=(-1)^{|\psi_1^{ab}||\psi_2^{ba}|}Tr_b(\psi_2^{ba}\psi_1^{ab})
\ee
which involves the traces in two generally different sectors $a$ and $b$. 

Also note that non-degeneracy of the metrics is equivalent with the 
following properties of the traces:
\bea
Tr(\phi_1\phi_2)~~~=0~~{\rm for~all~}~\phi_2~~&\Rightarrow& \phi_1=0~~\\
Tr_a(\psi^{ab}_1\psi^{ba}_2)=0 ~~{\rm for~all~}~\psi_2^{ba}&\Rightarrow& \psi_1^{ab}=0~~\\
Tr_a(\psi^{ab}_1\psi^{ba}_2)=0 ~~{\rm for~all~}~\psi_1^{ab}&\Rightarrow& \psi_2^{ba}=0~~.
\eea 

The maps $Tr, Tr_a$ 
need not agree with the linear algebra traces $tr_{{\cal H}_{ab}}$ on the 
spaces ${\cal H}_{ab}$. For the operators ${\hat \psi}_\alpha^{a}$, 
we have ${\hat \psi}^a_\alpha|\psi^a_\beta\rangle=
B(a)(|\psi_\alpha^a\rangle~,~|\psi_\beta^a\rangle)=B_{\alpha\beta}^\gamma(a)
|\psi_\gamma^a\rangle$ (where $B(a):=B(aaa)$). Hence:
\be
tr_{{\cal H}_a}({\hat \psi}_\alpha^a)=B_{\alpha\beta}^\beta(a)~~,
\ee
while:
\be
Tr_a(\psi_\alpha^a)=\rho_{\alpha0}(a)=\rho_{0\alpha}(a)~~.
\ee

In the topological case, the one-point functions $Tr(\phi_i)=\eta_{i0}=\eta_{0i}$ 
and $Tr_a(\psi^{a}_\alpha)=\rho_{\alpha 0}(a)=\rho_{0\alpha}(a)$ need not be zero 
for non-vanishing $i,\alpha$. Let us compare this situation with the case of 
boundary conformal field theories. For such systems, one 
has a unique state of conformal dimension $h=0$, the conformal vacuum. For a boundary 
state $\psi^{a}$ of nonzero dimension, conformal invariance requires 
$Tr_{\alpha}(\psi^{a})=\langle \psi^{aa}\rangle^{aa}_{0,1}=0$, so only the boundary vacua
$|0_a\rangle$ can 
have nontrivial one-point functions. In the topological case, 
this constraint cannot be applied. For the example of topological sigma
models, a 0-form observable transforms with unit weight under 
the diffeomorphism group, so noting can be concluded about its one-point function. 

In the conformal case, it is well-known that 
$\langle 1_a \rangle_{0,1}^{a,a}=Tr_a(|0_a\rangle)$ can have 
different values for different boundary labels $a$; in general, one cannot normalize these to have the same value. The same is true for topological 
theories. 

\subsection{The total boundary state space}

The properties of boundary correlators 
can be given a more transparent form. Consider the 
`total open state space' ${\cal H}_o:=\oplus_{ab}{{\cal H}_{ab}}$. 
On this space, we introduce a `total boundary product' $B$ defined 
as follows. If $\psi^{ab}_1\in {\cal H}_{ab}$ and $\psi^{bc}_2\in {\cal H}_{bc}$, 
then $B(\psi^{ab}_1,\psi^{bc}_2):=B(abc)(\psi^{ab}_1,\psi^{bc}_2)$. Then we 
extend 
$B$ in the obvious manner to a bilinear map defined for arbitrary elements $\psi_1,\psi_2$ of 
${\cal H}_o$.
We also define a total boundary topological metric $\rho$ on ${\cal H}_o$ through
the requirement of bilinearity and the condition that it reduces to 
$\rho(ab)$ on `pure' boundary states $\psi_1\in {\cal H}_{ab}, \psi_2
\in {\cal H}_{ba}$ (the metric is defined to be zero on states 
$\psi_1\in {\cal H}_{a_1b_1},\psi_2\in {\cal H}_{a_2b_2}$ which do not 
satisfy the constraints $a_1=b_2$ and $a_2=b_1$.). Non-degeneracy of all $\rho(ab)$
is equivalent with non-degeneracy of $\rho$. 
It is not hard to see that the two boundary sewing constraints are equivalent with 
the conditions:
\bea
B(B(\psi_1,\psi_2),\psi_3)&=&B(\psi_1,B(\psi_2,\psi_3))~~\\
\rho(B(\psi_1,\psi_2),\psi_3)&=&(-1)^{|\psi_1|(|\psi_2|+|\psi_3|)}\rho(B(\psi_2,\psi_3),\psi_1)
\\
&=&(-1)^{|\psi_3|(|\psi_1|+|\psi_2|)}\rho(B(\psi_3,\psi_1),\psi_2)~~.\nn
\eea
The first of these equations means that $B$ is associative, while the second
is the requirement that the `total boundary correlator'
$\langle\psi_1,\psi_2,\psi_3\rangle=Tr(\psi_1\psi_2\psi_3)$ (defined on ${\cal
H}_o$ in the obvious fashion\footnote{This correlator is defined to be zero on
`pure' boundary states $\psi_i\in {\cal H}_{a_ib_i}$ which fail to satisfy the
requirement $b_1=a_2, b_2=a_3,b_3=a_1$. The total boundary trace is defined
through $Tr:=\sum_{a}{Tr_a}$, and satisfies
$\rho(\psi_1,\psi_2)=Tr(B(\psi_1,\psi_2))$.})  be cyclically symmetric.

Moreover, it is easy to see that the `total boundary vacuum' 
$|0\rangle_o:=\sum_{a}{|0\rangle_a}$ is a unit for the total boundary product. 
We conclude that $({\cal H}_o,B,Tr)$ is a unital associative superalgebra 
over $\C$, endowed with a non-degenerate and graded-cyclic trace.
This structure is familiar from open string field theory \cite{Witten_SFT,Gaberdiel},
where it appears in a slightly different context. 
Since the boundary metric $\rho$ is invariant under the 
boundary product:
\be
\rho(B(\psi_1,\psi_2),\psi_3)=\rho(\psi_1,B(\psi_2,\psi_3))~~,
\ee
we conclude that 
$({\cal H}_o,B,\rho)$ is a {\em non-commutative Frobenius superalgebra}.

\subsection{Bulk-boundary crossing symmetry}

The constraints of figure 12 (d) and (e) can be formulated as follows:
\bea
B(abb)(\psi,e(b)(\phi))&=&(-1)^{|\phi||\psi|}B(aab)(e(a)(\phi),\psi)~~\label{sewing_de1}\\
B(aaa)(e(a)(\phi_1),e(a)(\phi_2))&=&e(a)(C(\phi_1,\phi_2))~~.\label{sewing_de2}
\eea

To simplify their analysis, define the total bulk boundary map $e:{\cal H}\rightarrow 
{\cal H}_o$ by $e:=\oplus_a{e(a)}$~~. The image of this map is contained 
in the `diagonal' subspace ${\cal H}_d:=\oplus_{a}{{\cal H}_{aa}}$. Then 
conditions (\ref{sewing_de1}) and (\ref{sewing_de2}) can be rewritten as:
\bea
B(\psi,e(\phi))&=&(-1)^{|\phi||\psi|}B(e(\phi),\psi)~~\label{sewing_de_reduced1}\\
B(e(\phi_1),e(\phi_2))&=&e(C(\phi_1,\phi_2))~~.\label{sewing_de_reduced2}
\eea
The second equation shows that $e$ is a morphism from the bulk ring 
$({\cal H},C)$ to the boundary ring $({\cal H}_o,B)$. This morphism preserves 
units, since $e(a)|0\rangle=|0\rangle_a$ (cf. (\ref{bb_vacua})).

If we define 
a multiplication ${\cal H}\times {\cal H}_o\rightarrow {\cal H}_o$ by:
\be
\phi\psi:=B(e(\phi),\psi)~~,
\ee
then the boundary ring $({\cal H}_o,B)$ becomes a superalgebra over the bulk ring 
$({\cal H},C)$. To finish the check of algebra properties, 
we have to show that:
\be
B(\psi_1,B(e(\phi),\psi_2))=(-1)^{|\phi||\psi_1|}
B(B(e(\phi),\psi_1),\psi_2)=(-1)^{|\phi||\psi_1|}B(e(\phi),B(\psi_1,\psi_2))
\ee
and: 
\be
B(e(C(\phi_1,\phi_2)),\psi)=B(e(\phi_1), B(e(\phi_2),\psi))~~.
\ee
These properties follow from the constraints (\ref{sewing_de_reduced1}) and 
(\ref{sewing_de_reduced2}) and from associativity of the boundary product $B$:
\bea
\!\!\!\!\!\!\!\!\!\!\!\!\!\!B(\psi_1,B(e(\phi),\psi_2))&=&B(B(\psi_1,e(\phi)),\psi_2)=
(-1)^{|\phi||\psi_1|}B(B(e(\phi),\psi_1),\psi_2)\nn\\
&=&(-1)^{|\phi||\psi_1|}B(e(\phi),B(\psi_1,\psi_2))~~
\eea
and:
\be
B(e(C(\phi_1,\phi_2)),\psi)=B(B(e(\phi_1),e(\phi_2)),\psi)=
B(e(\phi_1),B(e(\phi_2),\psi))~~.
\ee

We conclude the boundary ring $({\cal H}_o,B)$ has the structure of  unital
superalgebra over the supercommutative ring 
$({\cal H},C)$. 

\subsection{Bulk modular invariance}

It is easy to see that the condition of figure 12(b) 
does not give any further constraints 
on the bulk quantities $c$ and $\eta$ --- for a topological field theory, this reduces to 
the tautology $tr(e(.,\phi))=tr(e(.,\phi))$.

\subsection{Modular invariance on the cylinder}

In contrast with bulk modular invariance, modular invariance on the cylinder 
gives  
a non-trivial constraint on the theory. The reason is that the sewing condition
depicted in figure 12 (f) involves both closed and open cuts of the surface, 
and thus relates bulk and boundary data. This constraint can be written:
\be
\label{sewing_f}
\langle \psi_1,\psi_2\rangle^{ab}_{0,2}
=\langle f(a)(\psi_1),f(b)(\psi_2)\rangle_{0,0}
\ee
where we use subscripts $g,h$ to indicate the genus and number of boundaries
of the surfaces involved. 

The map $f(a):{\cal H}_a\rightarrow {\cal H}$ appearing in this equation 
is obtained by considering the surface of figure 4 (c) with its opposite 
orientation (see figure 13). 
In particular, the 
bulk state $|a\rangle_B:=f(a)|0_a\rangle$ is described by the surface of figure 14.
Geometrically, the surface of figure 14 
translates between {\em non-string} bounding circles $C_a$ and closed string 
circle boundaries $C$.

\hskip 1.0 in
\begin{center} 
\scalebox{1.0}{\epsfxsize=8cm \epsfbox{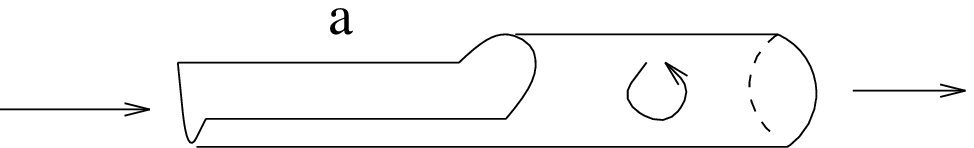}}
\end{center}
\begin{center} 
Figure  13. {\footnotesize The defining surface for the map $f$}
\end{center}

\hskip 1.0 in
\begin{center} 
\scalebox{1.0}{\epsfxsize=8cm \epsfbox{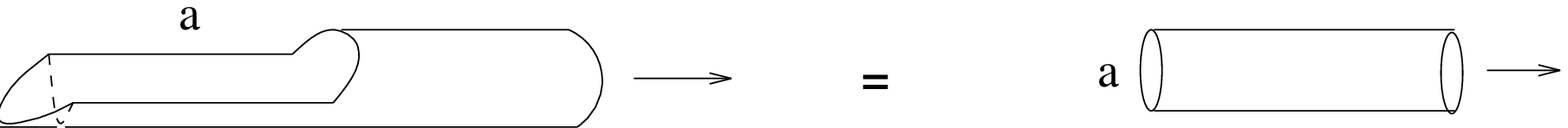}}
\end{center}
\begin{center} 
Figure  14. {\footnotesize The defining surface for the boundary state 
$|a\rangle_B=f(a)|0_a\rangle$. On the right, we have 
a cylinder which interpolates between a bounding circle $C_{a}$ (which, 
according to our definition, is not a `string boundary') and a closed string 
boundary $C$. As in the case of topological vacua, 
such a cylinder produces a closed string state ${\cal H}$ (=functional 
on the space of closed string configurations $V$ on the rightmost circle) 
upon considering the path integral with boundary condition $a$ at the left end.
Such a path integral contains no string state insertion on the left end, 
and hence this end can be allowed to be a non-string boundary.}
\end{center}

The map $f(a)$ is related to $e(a)$ as follows. If $\sigma$ is the 
surface of figure 4 (c) (with both string boundaries taken to be incoming), 
then the associated correlator 
$\langle\phi\psi\rangle^{a}_\sigma$ can be expressed in either of the following 
two ways:
\be
\label{adjointness}
\rho(a)(e(a)\phi,\psi)=\eta(\phi,f(a)\psi)~~.
\ee
Thus $e(a)$ and $f(a)$ are mutually adjoint 
with respect to the topological  metrics on their defining spaces.

One can rewrite relation (\ref{sewing_f}) as follows:
\be
\label{B_mod}
tr_{{\cal H}_{ab}}((-1)^F\Phi_{ab}
(\psi_1,\psi_2))=\eta(f(a)(\psi_1),f(b)(\psi_2))~~,
\ee
where $F$ is the `fermion number' (i.e. $F$ counts the degree of states), 
the map $\Phi_{ab}(\psi_1,\psi_2):{\cal H}_{ab}\rightarrow {\cal H}_{ab}$ is defined through:
\be
\label{cyl}
\psi\rightarrow B(abb)(B(aab)(\psi_1,\psi),\psi_2)~~,
\ee
and $tr_{{\cal H}_{ab}}$ is the usual trace in the vector 
space ${\cal H}_{ab}$.

To make this more specific, let us expand:
\bea
e(a)\phi_i=e_i^\alpha(a)\psi_\alpha^a~~\\
f(a)\psi_\alpha^a=f_\alpha^i(a)\phi_i~~.
\eea
Then (\ref{adjointness}) (applied to $\phi_i$ and $\psi_\alpha^a$) gives:
\be
\eta_{ij}f_\alpha^j(a)=e_i^\beta(a)\rho_{\beta \alpha}(a)~~.
\ee
Defining $f_{i\alpha}(a):=\eta_{ij}f_\alpha^j(a)$ and 
$e_{i\alpha}(a):=e^\beta_i(a)\rho_{\beta\alpha}(a)$, we obtain:
\be
\label{f}
f_{i\alpha}(a)=e_{i\alpha}(a) \Leftrightarrow f^i_\alpha(a)=\eta^{ij}\rho_{\beta\alpha}(a)
e^\beta_j(a)~~.
\ee
In matrix form:
\be
{\hat \eta}{\hat f}(a)={\hat e}^T(a){\hat \rho}(a)~~,
\ee
where ${\hat f}(a)_{i\alpha}:=f^{i}_\alpha(a)$, ${\hat e}(a)_{\alpha j}:=
e^\alpha_j(a)$ and ${\hat \rho}(a)_{\alpha\beta}:=\rho_{\alpha\beta}(a)$, 
${\hat \eta}_{ij}:=\eta_{ij}$.

Applying (\ref{B_mod}) to the states $\psi_1:=\psi_\alpha^{a}$ and 
$\psi_2:=\psi_\beta^{b}$ gives:
\be
\eta_{ij}f_\alpha^i(a)f_\beta^j(b)=(-1)^{|\sigma|}
B_{\alpha\sigma}^\gamma(aab)B_{\gamma\beta}^\sigma(abb)~~.
\ee
In this equation, we assumed that the basis $\psi^{ab}_\sigma$ is formed 
of states of definite degrees $|\psi^{ab}_\sigma|:=|\sigma|$.
Combining with expression (\ref{f}) for 
$f$, we can rewrite the constraint (\ref{B_mod})
as follows:
\be
\label{Cardy}
\eta^{ji}e_{i\alpha}(a)e_{j\beta}(b)=(-1)^{|\sigma|}
B_{\alpha\sigma}^\gamma(aab)B_{\gamma\beta}^\sigma(abb)~~.
\ee
This can be recognized as a topological version of 
the generalized\footnote{By generalized we mean that we apply the 
modular constraint of figure 12 (f) for arbitrary incoming 
states $|\psi_1\rangle$ and $|\psi_2\rangle$, 
and not only for $|\psi_1\rangle=|0_a\rangle$ 
and $|\psi_2\rangle=|0_b\rangle$ as is customary in the conformal field theory 
literature.} Cardy constraint (see below).

Finally, we formulate the bulk-boundary sewing constraints in terms of the 
total boundary space ${\cal H}_o$. We start by defining the total 
bulk-boundary map $e:=\oplus_{a}{e(a)}:{\cal H}\rightarrow {\cal H}_o$
and the total boundary-bulk map $f:=\sum_{a}{f(a)}:{\cal H}_{od}
\rightarrow {\cal H}$. Here ${\cal H}_{od}:=\oplus_{a}{{\cal H}_a}$ is 
the `diagonal' part of ${\cal H}_o$. 

Relation (\ref{adjointness})
implies that $e$ and $f$ are mutually adjoint with respect to the bulk metric and the 
total boundary metric:
\be
\eta(\phi,f(\psi))=\rho(e(\phi),\psi)~{\rm~for~all~~}\phi\in {\cal H}~
{\rm~and}~~\psi\in {\cal H}_o~~,
\ee
while the topological Cardy constraint becomes:
\be
\label{adjointness_summed}
\eta(f(\psi_1),f(\psi_2))=tr_{{\cal H}_o}((-1)^F\Phi(\psi_1,\psi_2))~~
{\rm for~all~~}\psi_1,\psi_2\in {\cal H}_o~~,
\ee
where $\Phi(\psi_1,\psi_2)$ is the endomorphism of ${\cal H}_o$ defined
through:
\be
\Phi(\psi_1,\psi_2)=\oplus_{ab}{\Phi_{ab}(\psi^{a}_1,\psi^{b}_2)}=
\left[\psi\in {\cal H}_o\rightarrow B(B(\psi_1,\psi),\psi_2)\right],
\ee
for all pairs of `diagonal' states $\psi_k=\oplus_{a}{\psi^{a}_k}\in 
{\cal H}_{o}$ ($k=1,2$), of components 
$\psi^{a}_k\in {\cal H}_{aa}$.

\paragraph{Boundary states}

In boundary conformal field 
theory, a relation similar with (\ref{sewing_f}) is used to define boundary 
states. Following this tradition, we call 
$|\psi\rangle^a_B:=f(a)|\psi\rangle\in {\cal H}$ the 
{\em topological boundary state} associated with the open string state 
$|\psi\rangle \in {\cal H}_a$ (this is an extension of the terminology 
used in the conformal case, as we shall see in a moment). 
Application of the sewing constraints allows one to reduce certain questions 
about the `diagonal' boundary sectors ${\cal H}_{aa}$ to problems for 
the associated boundary states. In particular, the correlator associated to
the surface $\sigma$ of figure 4 (c) can be written:
\be
\langle \phi,\psi\rangle_\sigma^{a}=\langle \phi,f(a)(\psi) \rangle_{0,0}=
\eta(\phi,\psi^a_B)=\langle e(a)(\phi),\psi\rangle^a_{0,1}=\rho(a)(e(a)(\phi),\psi)~~.
\ee

The adjointness relation (\ref{adjointness}) tells us that we can compute
tree-level bulk-boundary two-point functions $\langle \phi ,\psi^{a}\rangle$
either as $\langle e(a)(\phi) ,\psi^a\rangle^a_{0,1}$ (i.e. by pulling the
bulk state to the boundary and computing a correlator on the disk) or as
$\langle \phi ,f(a)(\psi^a)\rangle= \langle \phi ,\psi^a_B\rangle_{0,0}$,
i.e. by pushing the state $\psi$ to its bulk image (its associated boundary
state) and computing a correlation function on the sphere. Even though these
descriptions are equivalent at the level of two-point functions, such a
`duality' is not very powerful unless the maps $e$ and $f$ have appropriate
surjectivity/injectivity properties.

\paragraph{Limitations of the boundary state approach}

What is the precise 
power of the boundary state approach ? To answer this question, let us investigate
the properties of the maps $e(a)$ and $f(a)$. Consideration of figure 15
shows that $f(a)e(a)|\phi\rangle=C(f(a)|0_a\rangle,|\phi\rangle)=
C(|a\rangle_B,|\phi\rangle)$, where $|a\rangle_B:=f(a)|0_a\rangle$.  
The best we can hope for is to have 
$fe=\sum_{a}{f(a)e(a)}=id_{{\cal H}}$. In that case, $e=\oplus_a{e(a)}$
would be injective and $f:=\sum_a{f(a)}$ would be surjective. 
This can be achieved provided that 
$\sum_{a}{f(a)|0_a\rangle}=\sum_{a}{|a\rangle_B}=|0\rangle$, 
which is a sort of completeness constraint for the 
set of boundary conditions (D-branes) present in the theory. Thus, one 
can expect a simplification 
in the case when `all' topological D-branes have been 
included as a background, a situation which suggests that in a certain sense 
the bulk data can be recovered from knowledge of all possible boundary data.

\hskip 1.0 in
\begin{center} 
\scalebox{1.0}{\epsfxsize=8cm \epsfbox{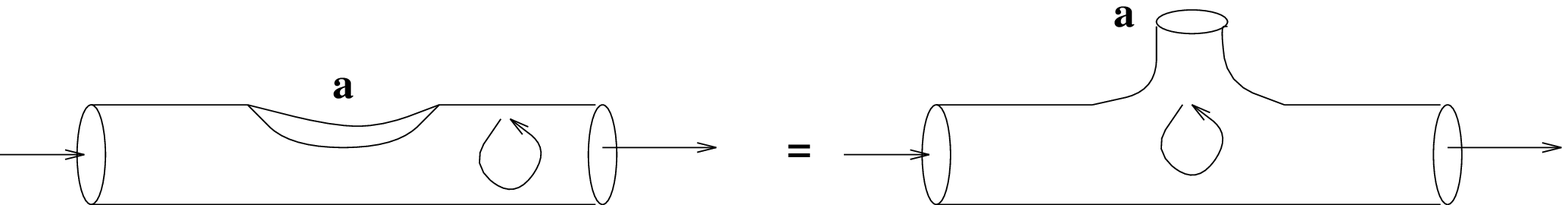}}
\end{center}
\begin{center} 
Figure  15. {\footnotesize Geometric description of the composition $f(a)e(a)$. This figure shows 
that $f(a)e(a)(\phi)=C(f(a)|0_a\rangle,\phi)$.}
\end{center}

\

Now consider the composition $e(b)f(a)$, which is shown in figure 16. 
This composition is 
also nontrivial, and nothing can be said about it from general considerations. 
Provided that $e(a)f(a)=id_{{\cal H}_a}$ in a given model, one could conclude that the 
boundary-bulk map is injective and the bulk-boundary map is surjective. Since 
the former is responsible for associating a boundary state to an open 
string state, this would assure us that the boundary state formalism gives
precise information on all open string states (i.e. we do not lose 
information on the spaces ${\cal H}_a$ when taking their images through
$f(a)$). Unfortunately, this is 
generally not the case, as one can see in the particular example of 
Calabi-Yau topological sigma models.

In general, the maps $e$ and $f$ 
are neither injective nor surjective. In particular, the boundary algebra 
$({\cal H}_o,B)$ will  generally have torsion as a module over the bulk algebra
$({\cal H},C)$. 

\hskip 1.0 in
\begin{center} 
\scalebox{1.0}{\epsfxsize=6cm \epsfbox{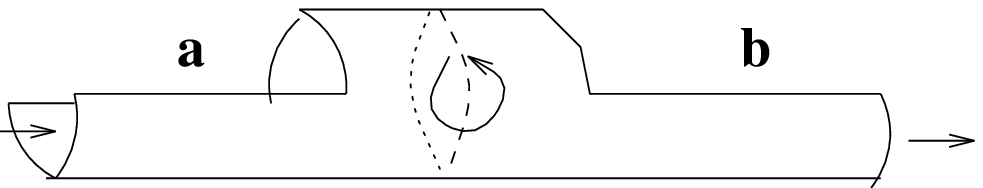}}
\end{center}
\begin{center} 
Figure  16. {\footnotesize Geometric description of the composition $e(b)f(a)$.}
\end{center}

\

We saw above that the map $e$ is a ring morphism from the bulk to the boundary 
algebra. It is natural to ask whether $f$ has a similar property. 
This is clearly not the case
\footnote{One can use relation (\ref{adjointness}) 
to show that $f$ is a ring morphism {\rm provided} that $ef=id$. Unfortunately, this 
is generally not true, as can be seen by considering the geometric interpretation 
of this composition (see below).}, 
since the quoted property of $e$ is a consequence 
of the constraint depicted in figure 12 (e), and no similar condition holds true 
for $f$. The closest candidate would be figure 12 
(d), which cannot be interpreted in such a manner since one cannot continuously 
pull the closed string tube in that figure along the open string strips in order
to produce two instances of the map $f$. 
\paragraph{On boundary states as D-branes}
Let us add a few remarks on the way boundary states are used in traditional boundary 
conformal field theory. In standard treatments, one is interested in computing 
couplings of the boundary {\em vacua} $|0_a\rangle$, and we can do the same in our topological 
field theories.  This amounts to considering only couplings of bulk states with the
{\em semiclassical} D-brane state associated with the boundary condition labeled by $a$. 
The associated boundary state $|a\rangle_B=|0_a\rangle_B\in {\cal H}$ is then called 
{\em the} boundary state defined by the D-brane $D_a$.  Then one writes the constraint
(\ref{B_mod}) for incoming states given by the boundary vacua 
$|0_a\rangle, |0_b\rangle$, in which case it reduces to:
\be
\eta(|a\rangle_B,|b\rangle_B)=\eta_{ij}f^i_0(a)f^j_0(b)=tr_{{\cal H}_{ab}}(-1)^F
~~.
\ee
This 
can be recognized as a topological version of the standard Cardy relation (note that $|a\rangle_B=\sum_i{f^i_0(a)|\phi_i\rangle}$).  The quantity in the right 
hand side is the Witten index of the boundary sector ${\cal H}_{ab}$.

It is more useful, however, 
to view the `quantum' D-brane $D_a$ as being defined by the entire 
open string state space ${\cal H}_{aa}$, since this includes all of the associated 
open string excitations. In fact, this is necessary for logical consistency since inclusion of 
D-branes should completely characterize our theory together with the bulk data. Since the 
boundary vacua $|0_a\rangle$ (and even more so the boundary states $|a\rangle_B$) 
{\em do not}
suffice to so characterize the theory, we must conclude that there is more information 
in D-brane physics than can be possibly encoded by the boundary vacua $|0_a\rangle$. It follows 
that the technique of boundary states represents at best a first step towards a complete 
characterization of boundary conformal/topological field theories. We should also note that
the literature on boundary states is almost exclusively concerned 
with the states $|a\rangle_B$, which 
encode information only about the boundary vacua.

Non-injectivity of $f(a)$ (if present) sets sharp limitations for the boundary state approach. 
In fact, this
formalism is further weakened by the following problem. Suppose that one is given a state 
$|\phi\rangle$ in the bulk space ${\cal H}$. One would like to know whether it corresponds
to an open string state $|\psi^a\rangle\in {\cal H}_{aa}$ for some $a$, 
and, if so, whether this state is unique.
This is precisely the approach followed by many studies of open conformal 
field theory through the boundary state formalism --- instead of trying to determine 
${\cal H}_{aa}$ directly, one tries to find a set of bulk states $|\phi\rangle$
which satisfy certain constraints (such as conformal invariance and 
Cardy's constraints in the case of rational conformal
field theories). This approach is especially useful in the case of abstract models (such as 
Gepner models) for which a worldsheet description of the associated boundary conditions
is difficult. 
Unfortunately, a solution of such constraints does {\em not} necessarily 
correspond to a true boundary state, since we are not assured that $|\phi\rangle$ lies 
in the image of any $f(a)$. To establish this for any given candidate, 
one needs to 
provide a full construction of a boundary extension of the bulk theory, and show that the 
candidate boundary state indeed is image through $f(a)$ of some open string state
(say, of the boundary vacuum $|0_a\rangle$). 
 
Moreover, the answer to this question could be highly ambiguous, since one does not know a priori what are the allowed boundary sectors $a$, 
nor does one know what the maps $f(a)$ are.
Even if one knew these maps, it is very possible to have $|\phi\rangle=
\sum_{a}{f(a)|\psi_a\rangle}$ for some nonzero 
states $|\psi^a\rangle\in {\cal H}_{aa}$, in which case 
$|\phi\rangle$ is not associated with any single state belonging to 
one of the spaces ${\cal H}_{aa}$. Since the 
full collection of possible boundary sectors $a$ is generally large, 
it is likely that 
this problem is quite widespread. 

The observations above raise some obvious questions about the extent to which 
recent work on open-closed Calabi-Yau compactifications \cite{Douglas_quintic,
  Douglas_follow} (which is largely based on the 
boundary state approach) gives us unambiguous information about the associated D-branes, 
and how seriously one can take the geometric interpretation of those states 
as type A/B D-branes wrapped over specific cycles in the large radius limit.

It is likely  that a complete understanding 
of open-closed extensions of a closed topological/conformal 
field theory must go beyond the boundary state approach. 
An attempt at classifying such extensions must explicitly include 
the freedom allowed by the choice of the maps 
$e(a)$, thus taking into consideration 
the bulk-boundary operator products as part of the extension 
problem. In this respect, progress 
has been made recently in \cite{Zuber}, where bulk-boundary products are discussed 
in certain rational conformal field theories.

\paragraph{What is an abstract `boundary condition'?}

In an abstract system, one lacks a direct construction of
the boundary theory through boundary conditions and boundary couplings in the action.
If one is interested in classifying all open-closed theories compatible with given
bulk data, one would like to have a conceptual definition of the `boundary part' 
of such a system. The traditional approach to this problem is to try to isolate 
the boundary data through use of the boundary state formalism, and to roughly 
identify the boundary states $|a\rangle_B$ with abstract `boundary configurations'.
As mentioned above, this approach encounters certain difficulties, the 
most obvious of which is that it does not take into account the bulk-boundary 
map $f(a)$. The latter is essentially `restriction to the boundary' in the 
standard case of topological sigma models
(with `geometric' boundary conditions). 
Therefore, its specification is crucial for any consistent definition of boundary 
data. I believe that the only generally meaningful procedure is to define 
`boundary data' as the entire system $({\cal H}_o,B,\rho,e)$. This reduces the 
task of classifying boundary theories associated to given bulk data to the problem 
of finding all non-commutative Frobenius superalgebras over $({\cal H},C,\eta)$ which 
obey the topological Cardy constraint. In such generality, this problem can be 
expected to have solutions which do not fit into a geometric 
`boundary condition' approach--for example, a topological sigma model or 
Landau-Ginzburg model at the conformal point could possess more open-closed
extensions than predicted by the classical boundary condition approach of
\cite{Witten_CS, LG_bcs}.  Whether such an abstract boundary extension has a
`classical' boundary condition interpretation or not is a model-dependent
question which is largely irrelevant if a complete solution of the problem is
known.

\subsection{Summary}

We showed that a (two-dimensional) topological open-closed field theory is 
equivalent with the following data:

(1) A Frobenius superalgebra $({\cal H},C,\eta)$ over 
the complex numbers. We recall that 
this is an associative,  supercommutative algebra $({\cal H},C)$ 
with unit $|0\rangle$, endowed with a graded-symmetric nondegenerate 
bilinear form $\eta$ (the bulk topological metric). This metric has 
a definite degree $deg(\eta)$ ( i.e. $\eta(\phi_1,\phi_2)=0$ unless 
$|\phi_1|+|\phi_2|=deg(\eta)$), which is model-dependent. 

(2) A collection of finite-dimensional super-vector spaces 
$({\cal H}_{ab})_{a,b\in \Lambda}$ indexed by a set $\Lambda$ (taken to be 
finite, for ease of exposition), together with degree zero bilinear maps 
$B(abc):{\cal H}_{ab}\times {\cal H}_{bc}\rightarrow {\cal H}_{ac}$ and 
bilinear maps 
$\rho(ab):{\cal H}_{ab}\times {\cal H}_{ba}\rightarrow \C$ of definite 
(but model-dependent) degree, with the following 
properties:

(2.1) $B(abc)(B(adb)(\psi_1,\psi_2),\psi_3)=
B(adc)(\psi_1,B(dbc)(\psi_2,\psi_3))$.

(2.2) $B(abb)(\psi,|0_b\rangle)=B(aab)(|0_a\rangle, \psi)=
\psi$ for some elements $|0_a\rangle \in {\cal H}_{a}$.

(2.3) $\rho(ab)$ are nondegenerate and satisfy 
the `graded-commutativity' relations:
\be
\rho(ab)(\psi_1,\psi_2)=(-1)^{|\psi_1||\psi_2|}
\rho(ba)(\psi_2,\psi_1)~~.
\ee
and the cyclicity property:
\bea
\rho(\psi_1,B(\psi_2,\psi_3))=(-1)^{|\psi_1|(|\psi_2|+|\psi_3|)}
\rho(\psi_2,B(\psi_3,\psi_1))=\\
=(-1)^{|\psi_3|(|\psi_1|+|\psi_2|)}
\rho(\psi_3,B(\psi_1,\psi_2))~~;\nn
\eea

(3) Degree zero linear maps $e(a):{\cal H}\rightarrow {\cal H}_a$ with the 
properties:

(3.1) $e(a)|0\rangle=|0_a\rangle$

(3.2) $B(aaa)(e(a)\phi_1,e(a)\phi_2)=e(a)(C(\phi_1,\phi_2))$

(3.3) $B(abb)(\psi,e(b)\phi)=(-1)^{|\phi||\psi|}B(aab)(e(a)\phi,\psi)$~~.

This data is such that:

(4) the topological Cardy constraint (\ref{B_mod}) is satisfied.

For the reader's convenience, let us explain how one can determine the basic data $\eta,\rho,
C,B,e$ from computations of topological correlators. It is convenient 
to chose the bases $\phi_i$ and $\psi^{ab}_\alpha$ of ${\cal H}$, ${\cal H}_{ab}$
to be homogeneous, i.e. such that $\phi_i$ are elements of definite 
degree $|i|$ and $\psi^{ab}_\alpha$ are elements of definite degree $|\alpha|$
\footnote{$|i|$ and $|\alpha|$ should not be confused with absolute values !}.
Since $\rho(ab)$ is nondegenerate and of definite degree, 
the spaces ${\cal H}_{ab}$ and ${\cal H}_{ba}$ are isomorphic 
(possibly after a shift of grading)  as graded vector spaces 
and hence the bases $\psi^{ab}_\alpha$ and 
$\psi^{ba}_\alpha$ are indexed by the same set of labels $\alpha$.
Raising/lowering indices with the bulk and boundary 
topological metrics, we define:
\bea
C_{ijk}&:=&C_{ij}^l\eta_{lk}~~\\
B_{\alpha\beta\gamma}(abc)&:=&B_{\alpha\beta}^\delta(abc)\rho_{\delta\gamma}(ac)~~\\
e_{i\alpha}(a)&:=&e_i^\beta(a)\rho_{\beta\alpha}(a)~~\\
f_{i\alpha}(a)&:=& \eta_{ij}f^{j}_\alpha(a)~~,
\eea
where we used the notations $\rho_{\alpha\beta}(a):=\rho_{\alpha\beta}(aa)$ etc.
One has:
\bea
C_{ijk}&=&\eta(C(\phi_i,\phi_j),\phi_k)=\langle \phi_i \phi_j \phi_k\rangle_{0,0}~~\\
B_{\alpha\beta\gamma}(abc)&=&\rho(ac)(B(\psi_\alpha^{ab},\psi_\beta^{bc}),
\psi_\gamma^{ca})=
\langle \psi_\alpha^{ab}\psi_\beta^{bc}\psi_\gamma^{ca}\rangle_{0,1}~~\\
e_{i\alpha}(a)~&=&\rho(a)(e(a)(\phi_i),\psi_\alpha^a)=
\langle \phi_i \psi_\alpha^a \rangle_{0,1}~~.
\eea
On the other hand, we have:
\be
\eta_{ij}=\langle \phi_i \phi_j\rangle_{0,0}~~,~~
\rho_{\alpha\beta}(ab)=\langle \psi^{ab}_\alpha\psi^{ba}_\beta\rangle_{0,1}~~.
\ee
Hence all relevant data can be determined by computing:

(1) The two and 3-point functions on the sphere $\eta_{ij}$ and $C_{ijk}$

(2) The boundary two and 3-point functions on the disk $\rho_{\alpha\beta}(ab)$ and 
$B_{\alpha\beta\gamma}(abc)$

(3) The bulk-boundary two-point function on the disk $e_{i\alpha}(a)=f_{i\alpha}(a)$

In coordinates, the constraints on this data are as follows:

(a) $\eta_{ij}$ and $\rho_{\alpha\beta}(ab)$ are non-degenerate and we have:
\bea
\eta_{ij}~~&=&(-1)^{|i||j|}\eta_{ji}~~\\
\rho_{\alpha\beta}(ab)&=&(-1)^{|\alpha||\beta|}\rho_{\beta\alpha}(ba)~~.
\eea 

(b) $C_{ijk}$ are graded symmetric:
\be
C_{ijk}=(-1)^{|i||j|}C_{jik}=(-1)^{(|i|+|j|)|k|}C_{kij}~{\rm~etc.}
\ee 
and form the structure constants of an associative algebra with unit. We 
can chose this unit $|0\rangle$ to be part of our basis: $\phi_0:=|0\rangle$.
In this case, we must have:
\bea
C_{0j}^k&=&\delta^k_j~~\\
C_{i0}^k&=&\delta^k_i~~.
\eea
Notice that $\eta_{jk}=C_{0jk}$ with this choice of basis.

(c) $B_{\alpha\beta\gamma}(abc)$ are graded cyclically symmetric:
\be
B_{\alpha\beta\gamma}(abc)=(-1)^{|\alpha|(|\beta|+|\gamma|)}
B_{\beta\gamma\alpha}(bca)=(-1)^{|\gamma|(|\alpha|+|\beta|)}B_{\gamma\alpha\beta}
(cab)~~
\ee
and satisfy the associativity property:
\be
B_{\alpha\beta}^\sigma(abc)B_{\sigma\gamma}^\delta(acd)=
B_{\alpha\sigma}^\delta(abd)B_{\beta\gamma}^\sigma(bcd)~~.
\ee  
They are also required to admit units $|0_a\rangle$ in the sense of Subsection 
3.6. Choosing $\psi^{aa}_0:=|0_a\rangle$, we can formulate this as the 
constraints:
\bea
B_{0\beta}^\gamma(aab)&=&\delta^\gamma_\beta~~\\
B_{\alpha0}^\gamma(abb)&=&\delta^\gamma_\alpha~~.
\eea
Notice that $\rho_{\beta\gamma}(ab)=B_{0\beta\gamma}(aab)$ with this choice of 
basis.

(d) $e_{i\alpha}$ induce a graded algebra structure of ${\cal H}_o$ over 
${\cal H}$, i.e. we have (see equations (\ref{sewing_de1}) and (\ref{sewing_de2})) :
\bea
B_{\alpha\beta}^\gamma(abb)e_i^\beta(b)~~~~~~~~&=&
(-1)^{|i||\alpha|}e_i^\beta(a)B_{ \beta\alpha }^\gamma(aab)~~\\
B_{\alpha\beta}^\gamma(aaa)e_i^\alpha(a)e_j^\beta(a)&=&~~~~~~e_k^\gamma(a)C_{ij}^k~~.
\eea

(e) The topological Cardy constraint (\ref{Cardy}) is satisfied.

As in closed topological field theory, the entire information is contained in tree-level
amplitudes
\footnote{This is of course {\em not} the case in topological {\em string} theory, i.e.
after coupling to topological gravity.}. Hence one can invert the argument and {\em define} 
an (on-shell) topological open-closed 
field theory to be given by such a structure. In particular, 
on-shell deformations of open-closed topological models 
are governed by the deformation theory of such objects.

\section{Decompositions and irreducible boundary theories}

We saw that the boundary and bulk-boundary 
sewing constraints admit a simplified form when expressed in terms of the 
total boundary state space. In this section, I analyze the structure of 
open-closed topological field theories from this point of view. Since the 
data summarized in Subsection 4.8. is quite complicated, it is useful to 
follow a formal approach and start with a few definitions. 

{\bf Definition 5.1.} A {\em bulk algebra} is a 
Frobenius superalgebra $({\cal H},C,\eta)$, whose unit we denote by $|0\rangle$. 
Remember that we take the invariant, nondegenerate metric $\eta$ to be 
graded-symmetric, as required by the equivariance axiom. We allow $\eta$ to
have a definite degree $n\in \Z_2$, which is model-dependent.

{\bf Definition 5.2.} A {\em boundary algebra} is a triple $({\cal H}_o,B,\rho)$
with the properties:

(1) ${\cal H}_o$ is a (finite-dimensional) complex vector space, endowed 
with a $\Z_2$-grading ${\cal H}_o={\cal H}_o^0\oplus {\cal H}^1$.

(2) $({\cal H}_o,B)$ is an associative superalgebra 
over the field of complex numbers. In particular,
the product $B:{\cal H}_o\times 
{\cal H}_o\rightarrow {\cal H}_o$ is bilinear with respect to complex 
multiplication and has degree zero. This algebra is endowed with a 
unit which we denote by $|0\rangle_o$. The product $B$ need not be 
graded-commutative.

(3) $\rho:{\cal H}_o\times {\cal H}_o\rightarrow \C$ is a nondegenerate  
bilinear form satisfying the properties:

(3.1.) graded symmetry:
\be
\rho(\psi_1,\psi_2)=(-1)^{|\psi_1||\psi_2|}\rho(\psi_2,\psi_1)~~.
\ee

(3.2) invariance:
\be
\rho(B(\psi_1,\psi_2),\psi_3)=\rho(\psi_1,B(\psi_2,\psi_3))
\ee

(3.3) graded cyclicity:
\be
\rho(\psi_1,B(\psi_2,\psi_3))=(-1)^{|\psi_1|(|\psi_2|+|\psi_3|)}
\rho(\psi_2,B(\psi_3,\psi_1))=\\
=(-1)^{|\psi_3|(|\psi_1|+|\psi_2|)}
\rho(\psi_3,B(\psi_1,\psi_2))~~.
\ee

{\bf Definition 5.3.} A {\em boundary extension} of a bulk algebra 
$({\cal H},C,\eta)$ is a boundary algebra $({\cal H}_o,B,\rho)$
together with a degree zero linear map $e:{\cal H}\rightarrow {\cal H}_o$, 
having the properties: 

(1) $e$ endows $({\cal H}_o,B)$ with the structure of a graded 
unital algebra over the bulk ring $({\cal H},C)$, i.e. we have:

(1.1) $e|0\rangle=|0\rangle_o$

(1.2) $B(e(\phi),\psi)=(-1)^{|\phi||\psi|}B(\psi,e(\phi))$

(1.3) $B(e(\phi_1),e(\phi_2))=e(C(\phi_1,\phi_2))$~~.

(2) The topological Cardy constraint:
\be
\eta(f(\psi_1),f(\psi_2))=tr[(-1)^F\Phi(\psi_1,\psi_2)]
\ee
is satisfied. Here $f$ is the adjoint of $e$ with respect to the metrics 
$\eta,\rho$:
\be
\label{adj_again}
\eta(\phi, f(\psi))=\rho(e(\phi), \psi)~~,
\ee
while $\Phi(\psi_1,\psi_2)$ is the endomorphism of ${\cal H}_o$ defined through:
\be
\psi\rightarrow B(B(\psi_1,\psi),\psi_2)~~.
\ee 

The module product ${\cal H}\times {\cal H}_o\rightarrow {\cal H}_o$ is 
defined in standard fashion, namely $\phi\psi:=B(e(\phi),\psi)$.
Note that complex multiplication in the boundary algebra is compatible 
with the extension map $e$ (since the later is linear). That is, the 
module structure of $({\cal H}_o,B)$ over $({\cal H},C)$ determined by $e$
takes complex multiplication in ${\cal H}$ into complex multiplication 
in ${\cal H}_o$:  
\be
\lambda \psi=B(e(\lambda |0\rangle),\psi)~~.
\ee

Boundary extensions correspond to open-closed topological field theories 
having a single boundary sector (i.e. when $\Lambda$ is a set with only one 
element). In Section 4, we expressed 
the sewing constraint in terms of the total boundary state space. In the 
language of the present section, what we did is to show that the axioms of 
open-closed topological field theory amount to the condition that the 
total boundary state space is an extension of the bulk algebra. To formulate 
this statement (and its converse) precisely, we need a few more mathematical 
definitions.

{\bf Definition 5.4} A {\em reduction} of a boundary extension 
$({\cal H}_o,B,\rho,e)$ is a (finite) direct sum decomposition 
${\cal H}_o=\oplus_{a,b\in \Lambda}{{\cal H}_{ab}}$ into (nonzero) 
vector spaces, with the properties:

(0) The decomposition is compatible with the grading, i.e. 
${\cal H}_{ab}=({\cal H}_{ab}\cap {\cal H}_o^{0})\oplus 
({\cal H}_{ab}\cap {\cal H}_o^{1})$
\footnote{This constraint is nontrivial, since it requires that any element 
$|\psi\rangle\in {\cal H}_{ab}$ be the sum of an even and an odd element of 
${\cal H}_o$ {\em both of which  belong to ${\cal H}_{ab}$.}}~~.

(1) $\rho$ is zero except when restricted to subspaces of the form 
${\cal H}_{ab}\times {\cal H}_{ba}$.

(2) The product $B$ is zero except when restricted to subspaces of the 
form ${\cal H}_{ab}\times {\cal H}_{bc}$, and takes such a space
into a subspace of ${\cal H}_{ac}$

(3) The map $e$ is `diagonal', i.e. its image $e({\cal H})$ is a subspace 
of $\oplus_{a\in \Lambda}{{\cal H}_{aa}}$.

Using condition (1), the requirement (2) can be reformulated as the statement 
that the triple correlator $\langle \psi_1 ,\psi_2 ,\psi_3\rangle=
\rho(\psi_1,B(\psi_2,\psi_3))$ equals zero except when restricted to subspaces 
of the form ${\cal H}_{ab}\times {\cal H}_{bc}\times {\cal H}_{ca}$. 
That is, the boundary topological metric and triple correlator must be 
`polygonal' in the sense described in figure 17. An extension 
of the bulk algebra will be called {\em reducible} if it admits a reduction 
and {\em irreducible} otherwise. A reduction will be called trivial 
if the set $\Lambda$ has only one element (in this case, the reduction 
`does nothing').

\hskip 1.0 in
\begin{center} 
\scalebox{1.0}{\epsfxsize=8cm \epsfbox{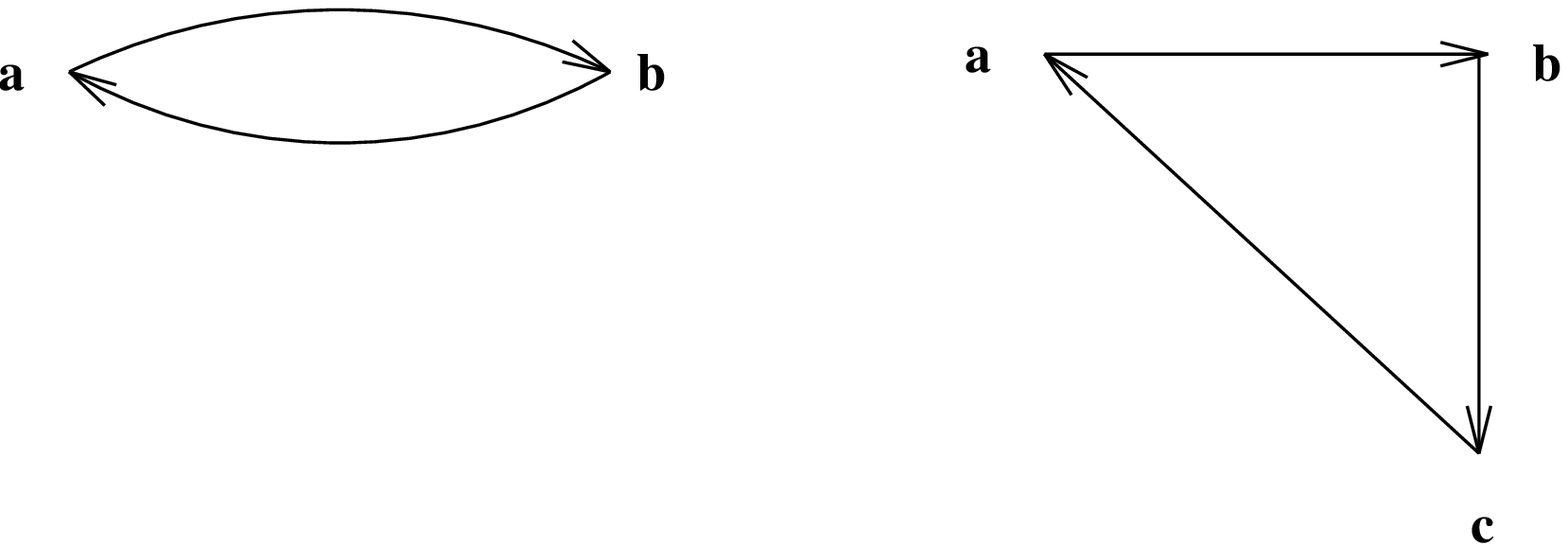}}
\end{center}
\begin{center} 
Figure  17. {\footnotesize Graphical representation of the 
decomposition conditions for the boundary metric and triple correlator. 
This description 
is related to the category-theoretic interpretation of 
subsection 5.2.}
\end{center}

The result of the sewing constraint analysis can now be formulated as follows:

{\bf Proposition 5.1} An open-closed extension of a closed topological field 
theory is equivalent with an extension $({\cal H}_o,B,\rho,e)$ 
of the bulk algebra $({\cal H},C,\eta)$, together with a (possibly trivial) 
reduction ${\cal H}_o=\oplus_{a,b\in \Lambda}{{\cal H}_{ab}}$. 

The fact that an open-closed topological field
theory defines such an extension and reduction was shown in Section 4. 
The converse (that an extension, together with a reduction, suffices to 
recover the data of Subsection 4.8) is rather obvious and I will not 
give all formal details here.
The only slightly subtle point is that the map $f:{\cal H}_o\rightarrow 
{\cal H}$ defined by the adjointness relation (\ref{adj_again})
has only `diagonal' components, i.e. is of the form 
$f=\sum_{a}{f(a)}$, with $f(a):=f|_{{\cal H}_{aa}}$. This follows upon writing
$f=\oplus_{ab}{f(ab)}$, with $f(ab)=f|_{{\cal H}_{ab}}$ and noticing that 
(\ref{adj_again}) implies
$\rho(a)(e(a)(\phi),\psi_{aa})=\eta(\phi,f(a)(\psi_{aa}))$ when applied 
to the diagonal components $\psi_{aa}$ of $\psi$. Hence 
(\ref{adj_again}) can be simplified to the form:
\be
\eta(\phi,\sum_{a\neq b}{f(ab)(\psi_{ab})})=0~~,
\ee
which must hold for all $\phi$. Since $\eta$ is non-degenerate, this 
requires $\sum_{a\neq b}{\psi_{ab}}=0$, so that $f(\psi)=
\sum_{a}{f(aa)(\psi_{aa})}$. Since $\psi$ is arbitrary, it follows that the 
map $f$ defined by the duality condition  (\ref{adj_again}) 
is diagonal. 

This result reveals a certain ambiguity in the construction of 
boundary sectors. In fact, one can have two realizations of the structure 
of Subsection 4.8, such that they both define the same total boundary 
extension $({\cal H}_o,B,\rho,e)$. Such realizations are indistinguishable from 
the physical point of view, and thus they should be identified.  

Consider the implications of this observation for the meaning of 
`topological D-branes'. The relevant problem is to give 
a precise formulation of what it means to determine `all' admissible 
boundary sectors $a$. 
Since any such sector $a$ defines an extension  
$({\cal H}_{aa},B(aaa),\rho(aa),e(a))$ of the bulk algebra, it is clear that 
each admissible label $a$ in Section 2 
must correspond to an object of the type described in 
Definition 5.3. However, such an extension need not be irreducible, 
and a reduction leads to a refinement of the admissible set of boundary labels.
It is clear that the meaningful procedure is to look for all 
{\em irreducible} boundary extensions. This allows us to give the following  

{\bf Definition 5.5} A topological D-brane (or `abstract boundary condition') 
is an {\em irreducible} 
boundary extension of the bulk algebra $({\cal H},C,\eta)$.

We can now give a precise 
formulation of the boundary extension problem for a given bulk 
topological field theory. Let $S$ be 
the set of (isomorphism classes of) irreducible boundary extensions
of a given bulk algebra (this set may be infinite). Assuming $S$ is known, 
a boundary extension of the bulk theory is constructed as 
follows. First, pick a (finite) subset $S_0$ of $S$, enumerated by a set 
of labels $\Lambda$, and let $({\cal H}_o^{(a)},B^{(a)},\rho^{(a)},
e^{(a)})\in S_0$ be the boundary extension  associated to $a \in \Lambda$.
Then boundary extensions of the given bulk theory, determined by this 
set of topological D-branes, correspond to structures of the type 
described in Subsection 4.8., based on the set of labels $\Lambda$ and
having the property that 
$({\cal H}_{aa},B(aaa),\rho(aa),e(a))=({\cal H}_o^{(a)},B^{(a)},\rho^{(a)},
e^{(a)})$ for all $a$ in $\Lambda$. This is a well-defined mathematical problem,
lying at the intersection between algebra and category theory 
(see the next subsection). 
To my knowledge, this program has not been carried out even for a 
single nontrivial example.

\subsection{Category-theoretic interpretation}

The presence of 
various boundary sectors gives a category-theoretic flavor to the boundary 
data of Subsection 4.8. Since this interpretation is quite obvious and 
not particularly deep, I will keep the following remarks short.
The categorical formulation arises by viewing the labels $a$ as objects and the 
open string states in ${\cal H}_{ba}$ as morphisms from $a$ to $b$: $Hom(a,b)={\cal H}_{ba}$.
Then one defines the composition of morphisms $Hom(a,b)\times Hom(b,c)\rightarrow Hom(a,c)$ 
by the boundary product $B(cba)$. This gives a category due 
to associativity of the boundary product $B$.
Note that we 
have $\Z_2$-gradings on the spaces ${\cal H}_{ba}$ and the composition
$B$ is a degree zero map. Thus, we have a $\Z_2${\em -graded 
category}. The boundary topological metric gives non-degenerate 
bilinear pairings $\rho(ba)$ between $Hom(a,b)$ and $Hom(b,a)$ which obey
the cyclicity constraint (\ref{cyclicity}). 
Restricting to degree zero morphisms ${\cal H}_{ba}^0$ gives 
an 'even' full sub-category. These are the fundamental objects behind the 
`categories of D-branes' recently considered in the literature 
\cite{Douglas_categories} from a space-time perspective. For the reader 
familiar with \cite{Douglas_categories}
\footnote{The authors of \cite{Douglas_categories} consider the 
Calabi-Yau B-model \cite{Witten_NLSM, Witten_mirror} 
in the presence of D-branes, some of which can be described 
as holomorphic bundles over the target space. For such D-branes, 
\cite{Douglas_categories}  considers 
the category of holomorphic bundles and bundle morphisms, and proposes a modified 
stability condition inspired by mirror symmetry arguments.}
, we mention that morphism compositions 
in our approach arise naturally from the {\em physical} 
boundary product, or equivalently
from triple boundary correlators. These physically-inspired compositions 
are not restricted to the even subspaces ${\cal H}_{ba}^0$.

If we view the boundary data in this fashion, then the bulk-boundary product acts as an 
outer multiplication on the morphism spaces $Hom(a,b)$ and endows them with a module 
structure over the Frobenius algebra $({\cal H},C)$. The topological 
Cardy condition 
(\ref{Cardy}) gives a supplementary constraint relating 
this outer multiplication and the boundary product $B$.

\section{Conclusions and directions for further research}

We gave a systematic derivation of the `on-shell' structure of 
two-dimensional topological field theories on oriented open-closed 
Riemann surfaces. The analysis of sewing constraints allowed us to
encode all information contained in such a theory in the 
well-defined algebraic structure summarized in Subsection 4.8. This 
mathematical object can be used as a {\em definition} of open-closed 
`on-shell' topological field theories, and can be taken as the starting 
point in the study of 
boundary extensions, as well as of 
the `on-shell' deformation 
problem. We drew attention to the central role of the bulk-boundary and 
boundary-bulk maps and gave a detailed analysis of the topological 
version (\ref{B_mod}) of the (generalized) Cardy constraint.

Consideration of an arbitrary number of `open string sectors' led us 
to the problem of decompositions of boundary extensions, 
a mathematical formulation of which 
was given in Subsection 5.1.  We also extracted a category structure
from the boundary products and proposed a precise definition of 
topological D-branes as irreducible boundary extensions. 

It is natural to ask about the off-shell counterpart of this framework in 
open-closed cohomological field theories. A correct treatment of this problem 
requires consideration of topological open-closed string field theory 
along the lines of \cite{Zwiebach}, generalized to the case of an arbitrary 
number of boundary sectors. In fact, open-closed string field theory holds the 
key to a physical understanding of recent mathematical work on homological 
mirror symmetry. 
The tree-level off-shell 
structure was recently discussed in \cite{Hofman}, under restrictive 
conditions on the boundary data 
\footnote{Reference \cite{Hofman} 
gives a 'partially off-shell' description of the tree-level algebraic 
structure in a boundary sector containing one D-brane, aimed at 
interpreting recent results of \cite{Kontsevich_Soibelman}.
Working at open 
string tree level, one does not consider the (off-shell) 
version of the topological Cardy constraint, which appears from a bulk-boundary sewing condition
on the cylinder but restricts tree level data. 
The sewing constraints considered 
in \cite{Hofman} realize the
vertex equations  of \cite{Zwiebach}.}.

Understanding homological mirror symmetry \cite{Kontsevich} 
from a string field theoretic point of view 
requires certain generalizations of the topological open string theories 
of \cite{Witten_CS}. In particular, one has to consider such systems in the presence 
of an arbitrary number of D-branes, which leads to boundary-condition-changing 
sectors. 

{\bf Note} After submitting this preprint, I was informed of 
unpublished work of G. Moore and G. Segal, who considered the same problem
independently, though apparently only for the ungraded case --- see Ref. \cite{Moore}
and the talk by G. Moore at http://online.kitp.ucsb.edu/online/mp01/moore2/.
I wish to thank G. Moore for bringing  his work to my attention.

\

{\bf Acknowledgments}
The author thanks Jae-Suk Park for many stimulating conversations 
and Martin Rocek for interest in his work. He also wishes to 
thank Sorin Popescu for mathematical observations.

\end{document}